\documentclass[twocolumn,final]{autart}
\usepackage{natbib}
\usepackage[utf8]{inputenc}
\usepackage{amsmath}
\usepackage{amsfonts}
\usepackage{amssymb}
\usepackage{mathrsfs}
\usepackage{paralist}
\usepackage{psfrag}
\usepackage{pstool}
\usepackage{enumitem}
\usepackage{mathtools}
\mathtoolsset{showonlyrefs=true}
\usepackage{color}
\usepackage{comment}
\usepackage{ifthen}
\newboolean{hub}
\newboolean{full}
\newcommand{\IfHub}[2]{\ifthenelse{\boolean{hub}}{{#1}}{{#2}}}
\newcommand{\IfFull}[2]{\ifthenelse{\boolean{full}}{{#1}}{{#2}}}
\makeatletter
\newcommand{\pushright}[1]{\ifmeasuring@#1\else\omit\hfill$\displaystyle#1$\fi\ignorespaces}
\newcommand{\pushleft}[1]{\ifmeasuring@#1\else\omit$\displaystyle#1$\hfill\fi\ignorespaces}
\makeatother
\newtheorem{theorem}{Theorem}

\newtheorem{definition}{Definition}
\newtheorem{proposition}{Proposition}
\newtheorem{lemma}{Lemma}
\newtheorem{corollary}{Corollary}
\newtheorem{assumption}{Assumption}

\newtheorem{example}{Example}

\newcommand{\R}[1]{\mathbb{R}^{#1}}
\newcommand{\N}{\mathbb{N}}
\newcommand{\sphere}[1]{\mathbb{S}^{#1}}
\newcommand{\sk}{S}
\newcommand{\SO}{\mathrm{SO}}

\newcommand{\ball}{\mathbb{B}}

\newcommand{\C}{C}
\let\crit\relax
\DeclareMathOperator{\crit}{crit}
\newcommand{\minus}{\backslash}

\newcommand{\Rnneg}{\mathbb{R}_{\geq 0}}
\newcommand{\Rpos}{\mathbb{R}_{> 0}}

\newcommand{\T}{\mathsf{T}}
\newcommand{\TS}[2][\mbox{}]{\T_{#1}\sphere {#2}}
\newcommand{\defneg}[1]{\mathbb{S}_{<0}^{#1}}
\newcommand{\defnegeq}[1]{\mathbb{S}_{\leq 0}^{#1}}
\newcommand{\defpos}[1]{\mathbb{S}_{>0}^{#1}}
\newcommand{\defposeq}[1]{\mathbb{S}_{\geq 0}^{#1}}
\DeclareMathOperator{\diag}{diag}
\DeclareMathOperator{\eulertodcm}{\tilde{\Rmat}}
\DeclareMathOperator{\sign}{sign}
\renewcommand{\d}[2]{\frac{d #1}{d #2}}
\renewcommand{\vec}[1]{\text{vec}\left( #1 \right)}

\DeclareMathOperator{\rge}{rge}
\newcommand{\tp}{^\top}
\newcommand{\inner}[2]{\left\langle #1,#2\right\rangle}
\newcommand{\norm}[2][\mbox{}]{\left|#2\right|_{#1}}
\DeclareMathOperator{\dom}{dom}
\newcommand{\grad}[2][\mbox{}]{\nabla_{#1}#2}
\newcommand{\pder}[2]{\frac{\partial #1}{\partial #2}}

\newcommand{\inv}{^{-1}}

\DeclareMathOperator{\trace}{trace}

\newcommand{\eigmin}{\lambda_{\text{min}}}
\newcommand{\eigmax}{\lambda_{\text{max}}}
\newcommand{\svmax}{\sigma_{\text{max}}}

\newcommand{\tto}{\rightrightarrows}
\newcommand{\pl}{^+}
\DeclareMathOperator*{\argmin}{\arg\,\min}
\DeclareMathOperator*{\argmax}{\arg\,\max}

\newcommand{\vleq}{\preceq}
\newcommand{\vl}{\prec}
\newcommand{\vgeq}{\succeq}
\newcommand{\vg}{\succ}
\newcommand{\projt}{{\downarrow_t}}

\newcommand{\Rmat}{R}
\newcommand{\x}{\times}
\newcommand{\eye}[1]{I_{#1}}

\newcommand{\ceq}{:=}

\newcommand{\D}[2][\mbox{}]{\Dmc_{#1}\left(#2\right)}
\newcommand{\sphtodcm}{\mathcal{R}}

\newcommand{\Amc}{\mathcal{A}}

\newcommand{\Dmc}{\mathcal{D}}
\newcommand{\Hmc}{\mathcal{H}}
\newcommand{\Lmc}{\mathcal{L}}

\newcommand{\Zmc}{\mathcal{Z}}

\newcommand{\Xmc}{\mathcal{X}}

\renewcommand{\tilde}[1]{\widetilde{#1}}
\newcommand{\bmtx}[1]{\begin{bmatrix}#1\end{bmatrix}}
\newcommand{\lbar}[1]{\underline{#1}}
\newcommand{\ubar}[1]{\overline{#1}}
\renewcommand{\hat}[1]{\widehat{#1}}
\newcommand{\pmtx}[1]{\begin{pmatrix}
#1
\end{pmatrix}}

\newcommand{\htd}{E}
\newcommand{\sol}{\phi}

\DeclareMathOperator{\Sat}{Sat}

\newcommand{\yset}[1][\mbox{}]{\mathcal{Q}_{#1}}
\newcommand{\gap}{\mu}
\newcommand{\sticky}{\mathcal{E}} % sticky set, all (x,x) \in \ss
\newcommand{\minVy}{\nu} % minimum of V over all y \in \yset
\newcommand{\yminV}{\varrho} % minimizer(s) of V(x,y) as a function of x

\newcommand{\PTSn}{\Pi} % projection onto the tangent space of Sn
\newcommand{\Vdom}{\mathcal{V}}
\newcommand{\gradx}{\nabla_{\tilde r}}
\newcommand{\yrbnd}{\gamma} % bound on <r,y>
\newcommand{\stickygap}{\Delta} % value of gap on sticky set, extended to all x \in \Sn
\newcommand{\maxVflow}{V^{*}} % max V over flow set
\renewcommand{\Xmc}{\mathcal{X}} % the reduced state space, \Sn \times \yset

\newcommand{\lboundr}{\alpha_1}
\newcommand{\uboundr}{\alpha_2}

\newcommand{\pe}{\tilde{p}}
\newcommand{\ve}{\tilde{v}}
\newcommand{\Td}{\rho}
\newcommand{\yaw}{\rho_2}
\newcommand{\yawd}{r_2^\star}
\newcommand{\sublevel}[1]{\Omega_{#1}}

\newcommand{\decayrate}{\lambda}
\newcommand{\pfc}[1][\mbox{}]{\mathscr{P}_{#1}}
\newcommand{\spf}[2][\mbox{}]{{#2}_{#1}}
\newcommand{\height}[1]{h_{#1}}

\newcommand{\Snsqr}{\sphere n \times \sphere n}
\newcommand{\geo}{c}
\newcommand{\I}{\mathcal{I}}
\newcommand{\length}[2][]{L^{#1}\left( #2 \right)}

\newcommand{\vfield}[1]{\Psi_{#1}}
\renewcommand{\tt}{\theta}
\newcommand{\ttset}{\Theta}
\newcommand{\paramset}{\Upsilon}
\newcommand{\grv}{g}
\newcommand{\wb}{\omega}
\newcommand{\trajset}{M_2\ball}

\newcommand{\xtwo}{\iota}
\newcommand{\uctrl}{\kappa_u}
\newcommand{\wctrl}[1]{\kappa_{#1}}

\newcommand{\lboundVone}{\lbar{\alpha}}
\newcommand{\uboundVone}{\ubar{\alpha}}
\newcommand{\ubounddVone}{\lambda_1}
\newcommand{\lboundVtwo}{\lbar{\alpha}_2}
\newcommand{\uboundVtwo}{\ubar{\alpha}_2}
\newcommand{\ubounddVtwo}{\lambda_2}
\newcommand{\domvfield}{\dom\vfield{}}

\newcommand{\adapt}{\nu^\star}

\newcommand{\Qlqr}{\hat{Q}}
\newcommand{\Rlqr}{\hat{R}}
\newcommand{\bound}{b}

\newcommand{\etatwo}{\hat\eta_2}

\newcommand{\ggrad}{^o}

\newtheorem{apro}{Proposition}[section]
\newtheorem{alem}{Lemma}[section]
\newtheorem{acor}{Corollary}[section]

\setboolean{hub}{false}
\setboolean{full}{true}

\begin{document}
\begin{frontmatter}
\title{Robust Global Exponential Stabilization on the $n$-Dimensional Sphere with Applications to Trajectory Tracking for Quadrotors\thanksref{footnoteinfo}} % Title, preferably not more 
                                                % than 10 words.

\thanks[footnoteinfo]{Pedro Casau is also with the Institute for Systems and Robotics at Instituto Superior Técnico, Universidade de Lisboa, Lisboa, Portugal.
C. Silvestre is with the Department of Electrical and Computer Engineering of the Faculty of Science and Technology of the University of Macau, Macau, China, on leave from Instituto Superior Técnico, Universidade de Lisboa, Lisboa, Portugal.
This work was partially supported by the projects MYRG2018-00198-FST and MYRG2016-00097-FST of the University of Macau; by the Macau Science and Technology, Development Fund under Grant FDCT/026/2017/A1 and by Fundação para a Ciência e a Tecnologia (FCT) through Project UID/EEA/50009/2019 and grant CEECIND/04652/2017. 
Research by R. G. Sanfelice partially supported by NSF Grants no. ECS-1710621 and
CNS-1544396, by AFOSR Grants no. FA9550-16-1-0015, FA9550-19-1-0053,
and FA9550-19-1-0169, and by CITRIS and the Banatao Institute at the
University of California.}

\author[UMAC]{Pedro Casau}\ead{pcasau@isr.ist.utl.pt},    % Add the 
\author{Christopher G. Mayhew}, %\ead{},  % (ead) as shown
\author[UCSC]{Ricardo G. Sanfelice}\ead{ricardo@ucsc.edu},               % e-mail address 
\author[UMAC]{Carlos Silvestre}\ead{cjs@isr.ist.utl.pt}

%\address[IST]{Department of Electrical and Computer Engineering, Laboratory for Robotics and Systems in Engineering and Science (LARSyS), Instituto Superior T\'{e}cnico, Universidade de Lisboa, 1049-001 Lisboa, Portugal}  % Please supply                                              
\address[UMAC]{Department of Electrical and Computer Engineering, Faculty of Science and Technology, University of Macau, Taipa, Macau, China.}        % here.
\address[UCSC]{Department of Computer Engineering, University of California, Santa Cruz, CA 95064, USA}             % full addresses

\begin{abstract}
In this paper, we design a hybrid controller that globally exponentially stabilizes a system evolving on the $n$-dimensional sphere, denoted by $\sphere n$. This hybrid controller is induced by a {``synergistic''} collection of potential functions on $\sphere n$. We propose a particular construction of this class of functions that generates flows along geodesics of the sphere, providing convergence to the desired reference with minimal path length.  We show that the proposed strategy is suitable to the exponential stabilization of a quadrotor vehicle.% and the behavior of the closed-loop system is illustrated by means of experimental results.
\end{abstract}

\end{frontmatter}

\section{Introduction}
\subsection{Motivation and Problem Statement}
In this paper, we design a hybrid controller for global exponential stabilization of a setpoint for a system evolving on the $n$-dimensional sphere, given by $\sphere n\ceq\{x\in\R{n+1}:x\tp x=1\}$. The dynamics of this system can be described by
\begin{equation}
\label{eq:kinematics}
\begin{aligned}
\dot{x}&=\PTSn(x) \omega & x&\in\sphere n,
\end{aligned}
\end{equation}
where $\omega\in\R{n+1}$ is the input and $\PTSn(x)=\eye{n}-x x\tp$ projects $\omega$ onto the tangent space to $\sphere n$ at $x$, given by $\TS[x]{n}\ceq\{z\in\R{n+1}:z\tp x=0\}.$ Even though there exist controllers that globally asymptotically stabilize a setpoint on the $n$-dimensional sphere (c.f.~\cite{mayhew_global_2013}) and others that globally exponentially stabilize a setpoint on the special orthogonal group of order 3 (c.f.~\cite{lee_global_2015} and~\cite{berkane_hybrid_2017}), to the best of our knowledge, the problem of global exponential stabilization of a setpoint in $\sphere n$ has not been addressed before, despite the fact that it has many meaningful applications, such as visual servoing (c.f.~\cite{triantafyllou_constrained_2018}); control of robotic manipulators (c.f.~\cite{chaturvedi_asymptotic_2009}); exoskeleton tracking (c.f.~\cite{brahmi_cartesian_2017}); multi-agent synchronization (c.f.~\cite{markdahl_almost_2018}); formation control (c.f.~\cite{zhao_bearing_2016}); rigid-body stabilization (c.f.~\cite{chaturvedi_rigid-body_2011}) and trajectory tracking for multi-rotor aerial vehicles (c.f.~\cite{mahony_multirotor_2012}), which we also explore in this paper.
To see how the proposed controller for global exponential stabilization on the $n$-dimensional sphere applies to {trajectory tracking for a multi-rotor aerial vehicle}, consider the following: the position dynamics of a multi-rotor vehicle can be described by 
\begin{equation}\label{eq:quadpv}
\begin{aligned}
\dot{p}&=v&
\dot{v}&=x u+\grv
\end{aligned}
\end{equation}
where $\grv$ is the acceleration of gravity, $p\in\R 3$ and $v\in\R 3$ denote the position and the velocity of the vehicle with respect to the inertial reference frame, {$u$ denotes the magnitude of the thrust} and $x\in\sphere 2$ denotes the direction of the thrust (c.f.~\cite{hamel_dynamic_2002}). Given a reference trajectory with acceleration $\ddot p_d$ and a control law $w:\R 3\to\R 3$ that exponentially stabilizes the double integrator, if the controller for $x$ exponentially stabilizes the commanded thrust direction, given by $\Td(\pe,\ve,\ddot p_d)\ceq\frac{w(\pe,\ve)-\grv+\ddot p_d}{\norm{w(\pe,\ve)-\grv+\ddot p_d}}$ for each $(\pe,\ve,\ddot p_d)\in\{(\pe,\ve,\ddot p_d)\in\R 9: w(\pe,\ve)-\grv+\ddot p_d\neq 0\}$ where $\pe$ and $\ve$ denote the position and velocity tracking errors, respectively, then exponential tracking is attained. This, however, cannot be achieved through continuous feedback because it is not possible to globally asymptotically stabilize a given setpoint on a compact manifold by means of continuous feedback (c.f.~\cite{bhat_topological_2000}). Moreover, if a dynamical system cannot be globally asymptotically stabilized through continuous feedback, it cannot be robustly stabilized by discontinuous feedback either, as shown in~\cite{mayhew_topological_2011}. Overcoming these limitations is particularly important for multi-rotor aerial vehicles due to their popularity and the wide range of applications in which they are used, such as surveillance, tracking, search and rescue, infrastructure inspection, agriculture {and} disaster mitigation (see e.g~\cite{fink_planning_2011},~\cite{mellinger_trajectory_2012},~\cite{jiang_inverse_2013},~\cite{augugliaro_dance_2013},~\cite{lupashin_platform_2014},~\cite{floreano_science_2015} and~\cite{liang_dynamics_2018}). In this paper, we present a hybrid controller that not only provides global exponential stability, but also confers a quantifiable robustness margin to perturbations.

{Recent} developments on hybrid control theory overcame the topological obstructions to global stabilization on compact manifolds with the introduction of synergistic potential functions in~\cite{mayhew_global_2010}. A potential function on a compact manifold is a continuously differentiable function that is positive definite relative to a given point, thus it induces a gradient vector field which asymptotically stabilizes the given reference from every initial condition except a set of measure zero. To see this, consider the closed-loop system resulting from the interconnection of the gradient-based feedback law of a potential function $\height{r}$ that is positive definite relative to $r\in \sphere n$ and~\eqref{eq:kinematics}, given by:
\begin{equation}
\label{eq:cfeed}
\begin{aligned}
\dot x=-\PTSn(x)\grad \height{r}(x).
\end{aligned}
\end{equation}
{The equilibrium points of~\eqref{eq:cfeed}} are the critical points of $\height{r}(x)$, denoted by
\begin{equation}
\label{eq:critV}
\begin{aligned}
\crit\height{r}\ceq\{x\in\sphere n: \PTSn(x)\grad\height{r}(x)=0\},
\end{aligned}
\end{equation}
{which correspond to the set of points $x$ where $\grad \height{r}(x)$ is orthogonal to $\TS[x]{n}$. Since the set~\eqref{eq:critV} includes} the maximum and the minimum of $\height{r}$ on $\sphere n$, it follows that $r\in\sphere n$ is not globally asymptotically stable for~\eqref{eq:cfeed}. On the other hand, synergistic potential functions are collections of potential functions that {enable a controller to achieve robust global asymptotic stabilization of} the given setpoint, because, at the undesired equilibria, there exists another function in the collection with a lower value that we can switch to.

{The concept of synergistic potential functions has been introduced to address the problem of stabilizing a three-dimensional pendulum in~\cite{mayhew_global_2010} and later used in full attitude stabilization in~\cite{mayhew_synergistic_2013},~\cite{lee_global_2015}; attitude synchronization~\cite{mayhew_quaternion-based_2012}; partial attitude stabilization (c.f.~\cite{mayhew_global_2013}), and stabilization by hybrid backstepping (c.f.~\cite{Mayhew2011}). More recently, the interest in this control technique has spawned the design of new synergistic potential functions on $\SO(3)$, such as the ones by~\cite{berkane_design_2015},~\cite{berkane_construction_tac2016} and~\cite{berkane_hybrid_2017}.} It was shown by the authors in~\cite{casau_robust_2015} that synergistic potential functions can be used for global asymptotic stabilization of a reference trajectory for a multi-rotor aerial vehicle, but the extent to which exponential stabilization is possible was not addressed.

\subsection{Contributions}

The contributions in this paper are as follows. \emph{Extension to the concept of synergistic potential functions:} in Section~\ref{sec:synergistic}, we show that the existence of a centrally synergistic potential function $V:\sphere n\x\yset\to\Rnneg${, where $\yset$ is a compact set,} induces a gradient-based control law that renders 
\begin{equation}
\label{eq:Amc}
\begin{aligned}
\Amc\ceq\{(x,y)\in\sphere n\x\yset:x=r\}
\end{aligned}
\end{equation}
globally asymptotically stable for the closed-loop system. This nomenclature is inherited from~\cite{mayhew_global_2013}, where synergistic potential functions satisfying $V(r,y)=0$ for all $y\in\yset$ are said to be central because they share are a common minimum at $r$. In this paper, we extend the previous notion of centrally synergistic potential functions, because we consider that $\yset$ is compact rather than finite, which adds flexibility to the design of synergistic potential functions. The proposed controller is significantly different from the one in~\cite{mayhew_synergistic_2013} because the focus is on global exponential stabilization of $\sphere n$ rather than $\SO(3)$ and it further expands the work in~\cite{mayhew_global_2010} from global asymptotic stabilization on $\sphere 2$ to global exponential stabilization on $\sphere n$. Interestingly, the controller that we propose may be used for global stabilization on $\sphere 3$ which is the universal cover of $\SO(3)$. However, the resulting controller would be more complex than that of~\cite{mayhew_quaternion-based_2011} if used for rigid-body stabilization, because it would not take advantage of the fact that $\sphere 3$ is a double cover of $\SO(3)$.
\emph{Global exponential stability of a setpoint on $\sphere n$:} in Theorem~\ref{theorem:expStability}, we show that, if $V$ is bounded from above and below by a polynomial function of the distance to $\Amc$ and if $V$ converges exponentially fast to $0$, then $\Amc$ is globally exponentially stable for the closed-loop system, in the sense that, for all initial conditions, the state of the system converges exponentially to $\Amc$. This is different from the controllers proposed in~\cite{lee_global_2015} and~\cite{berkane_hybrid_2017}, because these address the problem of global exponential stabilization on $\SO(3)$ rather than $\sphere n$.
\emph{Optimal switching:} in Section~\ref{sec:construct}, we construct a synergistic potential function on the $n$-dimensional sphere that meets the requirements for exponential stability and has an optimal switching law, in the sense that it guarantees that solutions to the closed-loop system follow geodesics whenever a jump of the hybrid controller is triggered.
\emph{Robustness to perturbations:} the proposed hybrid controller satisfies the so-called hybrid basic conditions, therefore it is endowed with nominal robustness properties that are outlined in~\cite{goebel_hybrid_2012}. In addition, the switching rule introduces a hysteresis gap that prevents chattering.
\emph{Saturated thrust feedback for exponential tracking of a reference trajectory for a multi-rotor aerial vehicle:} in Section~\ref{sec:quad}, we employ the hybrid controller for global exponential stabilization on $\sphere n$ in trajectory tracking for a multi-rotor aerial vehicle. We show that, for each compact set of initial position and velocity tracking errors and for all initial orientations, the reference trajectory is exponentially stable. This is particularly difficult, because in addition to the topological constraints to global attitude stablilization, multirotor aerial vehicles are subject to underactuation constraints that prevent stabilization of the vehicle when the commanded thrust is zero (c.f.~\cite{lizarraga_osbtructions_2004}). This issue has been widely acknowledged but mostly overlooked due to its singular nature. For example, in~\cite{lee_nonlinear_2013} this flight condition is assumed to not occur and in~\cite{hua_control_2009} the controller is turned off when the commanded thrust approaches zero. Similarly to the work of~\cite{hua_control_2015}, we assume that the reference trajectory does not lead to a situation where the commanded thrust is zero, but, unlike the aforementioned approach, we explicitly build this limitation into the control design procedure so to achieve semi-global exponential stability with respect to the position and velocity errors. {A video of experimental runs using this controller can be found in~\cite{Casau2017}.}

The paper is organized as follows: in Section~\ref{sec:not}, we present the notation and the framework of hybrid dynamical systems that is used in this paper. In Section~\ref{sec:synergistic}, we develop the notions of synergistic potential potential functions on the $n$-dimensional sphere. In Section~\ref{sec:quad}, we apply the given controllers to the tracking of a reference trajectory for a vectored-thrust vehicle, and in Section~\ref{sec:conclusions}, we present the conclusions of this work. {In~\cite{Casau2015} we reported the results that are presented in Section~\ref{sec:synergistic} without the proofs. In~\cite{casau_exponential_2016}, we presented some preliminary results on work reported in Section~\ref{sec:quad} without the constructive controller synthesis that is presented in this paper for the quadrotor application.}

\section{Preliminaries}\label{sec:not}

The symbol $\R{}$ denotes the set of real numbers, $\N$ denotes the set of natural numbers and zero, $\Rnneg\ceq\{x\in\R{}: x\geq 0\}$, $\R n$ denotes the $n$-dimensional Euclidean space equipped with the norm $\norm{x}\ceq\sqrt{\inner{x}{x}}$ for each $x\in\R n$, where $\inner{u}{v}\ceq u\tp v$ for each $u,v\in\R n$. The canonical basis for $\R n$ is denoted by $\{e_i\}_{1\leq i \leq n}\subset\R n$ and $c+r\ball\ceq\{x\in\R n:\norm{x-c}\leq r\}$. If a real symmetric matrix $A\in\R{n\x n}$ is positive (negative) definite, we write $A\in\defpos{n}$ ($A\in\defneg{n}$) or $A\vg 0$ ($A\vl 0$) if the dimensions can be inferred from context. If a real symmetric matrix $A\in\R{n\x n}$ is positive (negative) semidefinite, we write $A\in\defposeq{n}$ ($A\in\defnegeq{n}$) or $A\vgeq 0$ ($A\vleq 0$) if the dimensions can be inferred from context.
Given $A\in\R{m\x n}$, $\svmax(A)$ denotes the maximum singular value of $A$. The gradient of a continuously differentiable function $V:\R n\to\R{}$ is given by $\grad V(x)\ceq\bmtx{\pder{V}{x_1}(x) & \ldots & \pder{V}{x_n}(x)}$ for each $x=(x_1,\ldots,x_n)\equiv\bmtx{x_1 & \ldots & x_n}\tp\in\R n$. {The derivative of a differentiable matrix function with matrix arguments $F:\R{m\x n}\to\R{k\x \ell}$ is given by $
\D[X]{F(X)}\ceq\partial\vec{F(X)}/\partial\vec{X}\tp$ for each $X\in\R{m\x n}$, where $\vec{X}\ceq\bmtx{e_1\tp X\tp & \ldots & e_n\tp X\tp}\tp$.} {The domain of a set-valued mapping $M:\R n\tto \R m$ is given by $\dom  M\ceq\{x\in\R n: M(x)\neq\emptyset\}.$ The range of $M$ is the set $\rge M\ceq\{y\in\R n: \exists x\in\R m\text{ such that }y\in M(x)\}.$}
A hybrid system $\Hmc$ defined on $\R n$ can be represented by 
{%
\begin{equation}
\label{eq:hybrid}
\Hmc:\left\{
\begin{aligned}
\dot x &\in F(x) & x&\in C \\
x\pl &\in G(x) & x&\in D 
\end{aligned}
\right .
\end{equation}%
}
where $C\subset\R n$ is the flow set, $F:\R n\tto \R n$ is the flow map, $D\subset\R n$ and $G:\R n\tto \R n$ with $D\subset \dom G$ is the jump map, as defined in~\cite{goebel_hybrid_2012}.  The maps $F$ and $G$ are set-valued maps satisfying $C\subset \dom F$ and $D\subset\dom G$, {respectively.} 
Loosely speaking, solutions to hybrid systems are hybrid arcs that are compatible with the data of the hybrid system, i.e., functions $(t,j)\mapsto\sol(t,j)$ defined on a hybrid time domain $\htd\subset\Rnneg\x\N_0$ that satisfy $\dot\sol(t,j)\in F(\sol(t,j))$ for almost all $t\geq 0$ and $\sol(t,j)\in D$, $\sol(t,j+1)\in G(\sol(t,j))$ for every $\sol(t,j)$ and $\sol(t,j+1)$ belonging to $\dom\sol$. We say that a solution $\sol$ to~\eqref{eq:hybrid} is maximal if it cannot be extended, it is complete if its domain is unbounded, it is discrete if $\dom\phi\subset\{0\}\x\N$ and it is Zeno if it is complete and $\sup_t\dom\phi<+\infty$ (for a more rigorous description of solutions to hybrid systems see~\cite{goebel_hybrid_2012}). 
Under the assumption of completeness of maximal solution to $\Hmc$, a compact set $\Amc$ is said to be:
 stable for $\Hmc$, if for each $\epsilon>0$ there exists $\delta>0$ such that for each solution $\sol$ to $\Hmc$ with $\norm[\Amc]{\sol(0,0)}\leq\delta$ satisfies $\norm[\Amc]{\sol(t,j)}\leq\epsilon$ for each $(t,j)\in\dom\sol$;
attractive for $\Hmc$ if $\lim_{t+j\rightarrow\infty}\norm[\Amc]{\sol(t,j)}=0$,
where $\norm[\Amc]{x}:=\min_{y\in\Amc}\norm{x-y}$ and $\norm{x}\ceq \sqrt{\inner{x}{x}}$ for each $x\in\R{n}$. A set $\Amc$ is globally asymptotically stable for $\Hmc$ if it is both globally attractive and globally stable for $\Hmc$. We say that a compact set $\Amc\subset\R n$ is exponentially stable {in the $t$-direction} from $U$ if there exists $k,\lambda>0$ such that $\sup_t\dom\phi = \infty$ and $\norm[\Amc]{\phi(t,j)}\leq k\exp(-\lambda t)\norm[\Amc]{\phi(0,0)}$ for each maximal solution $\sol$ to the hybrid system with $\sol(0,0)\in U$ and $(t,j)\in\dom\phi$.

\section{Synergistic Potential Functions on $\sphere n$}\label{sec:synergistic}

\begin{comment}
.
An example of a potential function on $\sphere n$ is the height function relative to $r\in\sphere n$ given by
\begin{equation}\label{eqn:height}
\height{r}(x)\ceq 1-r\tp  x,
\end{equation}
for each $x\in\sphere n$, satisfying the following properties.

\begin{lemma}\label{lem:height}
Given $r \in \sphere n$, $\height{r}$ defined in \eqref{eqn:height} satisfies
\begin{align}
\label{eqn:gradh}
\grad \height{r}(x) &= -r \\
\label{eqn:crith}
\crit \height{r}  &= \{-r,r\}
\end{align}
and thus,
\begin{align}
\label{eqn:heightcrits}
\argmin_{x \in \sphere n} \height{r}(x) &= r  & \argmax_{x \in \sphere n} \height{r}(x) &= -r  \\
\label{eqn:heightminmax}
\height{r}(r) &= 0 & \height{r}(-r) &= 2.
\end{align}
In particular, $\height{r}$ is positive definite on $\sphere n$ relative to $r$.
\end{lemma}
\begin{pf}
Clearly, \eqref{eqn:gradh} follows directly from the definition of $\height{r}$ in \eqref{eqn:height} and basic rules of calculus. Then, if $x \in \crit \height{r}$, it follows that $\PTSn(x)r = 0$. In other words, $r$ has no component orthogonal to $x$ and thus $x \in \{-r,r\}$, which proves \eqref{eqn:crith}. Since $\height{r}$ is differentiable, the maximum and minimum values of $\height{r}$ are found at its critical points. Since there are only two critical points, \eqref{eqn:heightcrits} follows directly from \eqref{eqn:crith} and \eqref{eqn:heightminmax} (both of which follow by direct substitution). 
\qed\end{pf}
\end{comment}

In this section, we design a hybrid controller that \emph{globally} exponentially stabilizes a given reference {$r\in\sphere n\ceq\{x\in\R{n+1}:x\tp x=1\}$} for the system~\eqref{eq:kinematics} using the notion of centrally synergistic potential functions given next. 

\begin{definition}
A function $h_r:\sphere n\to \R{}$ is said to be a potential function on $\sphere n$ relative to $r$ if it is continuously differentiable and positive definite relative to $r\in\sphere n$, i.e., $h_r(x)\geq 0$ for each $x\in\sphere n$ and $h_r(x)=0$ if and only if $x=r$. We denote the collection of potential functions on $\sphere n$ relative to $r$ by $\pfc[r]$.\hfill$\square$
\end{definition}

\begin{definition}\label{dfn:synergisticFunc}
Given a compact set $\yset$, a function $V:\sphere n\x\yset\to\R{}$ is said to be a centrally synergistic potential function on $\sphere n$ relative to $r$ if the following hold:
\begin{inparaenum}
\item For each $y\in\yset$, $V^y$ is a potential function on $\sphere n$ relative to $r$, where $V^y(x)\ceq V(x,y)$ for each $x\in\sphere n$;
\item There exists $\delta>0$ such that 
\begin{equation}
 \label{eq:mu}
 \begin{aligned}
 \gap(x,y)\ceq V(x,y)-\minVy(x) >\delta
 \end{aligned}
 \end{equation} 
 for each \IfFull{$(x,y)\in\sticky(V),$ where%
 \begin{equation}
 \label{eq:stickyV}
 \begin{aligned}
\sticky(V)\ceq\{(x,y)\in\sphere  n\x\yset:\PTSn(x)\grad V^{y}(x)=0,x\neq r\}
 \end{aligned}
 \end{equation}
 and}{$(x,y)\in\sticky(V)\ceq\{(x,y)\in\sphere  n\x\yset:\PTSn(x)\grad V^{y}(x)=0,x\neq r\}$, where} 
{\begin{equation}
\label{eq:minVy}
\minVy(x)\ceq \min_{\bar y\in \yset} V(x,\bar y).
\end{equation}}
\end{inparaenum}
\noindent We denote the collection of centrally synergistic potential functions on $\sphere n$ relative to $r$ by $\spf[r]{\yset}$.\hfill$\square$
\end{definition}
Given a centrally synergistic potential function $V$, the function $(x,y)\to\gap(x,y)$ is referred to as the \emph{synergy gap} of $V$ at $(x,y)$. The synergy gap measures the difference between the current value of the function and all other potential functions in the collection. If~\eqref{eq:mu} holds, we say that $V$ has \emph{synergy gap exceeding $\delta$}. 
Given $V\in\spf[r]{\yset}$, we define the hybrid controller with output $\omega\in\R{n+1}$, state $y\in\yset$ and input $x\in\sphere n$ as follows:
\begin{subequations}
\label{eq:controller}
\mathtoolsset{showonlyrefs=false}
\begin{align}
\begin{rcases}
\omega = -\grad{V^y(x)}\\ 
\dot{y} = 0
\end{rcases}\  &(x,y)\in C\ceq\{(x,y)\in\Xmc:\gap(x,y)\leq\delta\}\\ 
y^+ \in\yminV(x)\phantom{-\grad{V^y}}
 &(x,y) \in D\ceq\{(x,y)\in\Xmc:\gap(x,y)\geq\delta\},
\end{align}
\end{subequations}
where $\Xmc:=\sphere n\x\yset$ and 
\begin{equation}
\begin{aligned}
\label{eq:yminV}\yminV(x)&:=\argmin_{\bar y\in \yset}V(x,\bar y).
\end{aligned}
\end{equation}
 The interconnection between~\eqref{eq:kinematics} and~\eqref{eq:controller} is represented by the closed-loop hybrid system $\Hmc:=(C,F,D,G)$, given by
\begin{equation}
\label{eq:closed_loop}
\begin{aligned}
\pmtx{\dot{x}\\ \dot{y}}&= F(x,y) = 
\pmtx{-\PTSn(x)\grad{V^y(x)}\\ 
 0} & (x,y)&\in C \\
\pmtx{x^+\\ y^+}&\in G(x,y)=\pmtx{x\\ \yminV(x)} & (x,y)&\in D.
\end{aligned}
\end{equation}
The parameter $\delta$ that is used in the hybrid controller~\eqref{eq:controller} is central to the construction of synergistic potential functions and it defines a hysteresis gap which prevents chattering. 
{This construction is similar across many earlier works on synergistic hybrid feedback (see e.g.,~\cite{mayhew_global_2013},~\cite{mayhew_synergistic_2013}) and it guarantees global asymptotic stability of $x=r$ for the hybrid system~\eqref{eq:closed_loop}, as shown next.}

\begin{theorem}\label{theorem:gas}
Given $r\in\sphere n$ and a compact set $\yset$, if there exists $\delta>0$ such that $V\in\spf[r]{\yset}$ has synergy gap exceeding $\delta$, then the set $\Amc$ in~\eqref{eq:Amc} is globally asymptotically stable for the hybrid system $\Hmc$ in~\eqref{eq:closed_loop}.
\end{theorem}
\begin{pf}
%To show that $\Amc$ is globally pre-asymptotically stable for~\eqref{eq:closed_loop}, we start by demonstrating that~\eqref{eq:closed_loop} satisfies~\cite[Assumption~6.5]{goebel_hybrid_2012}. It follows from Proposition~\ref{prop:muIsNice} that $\gap$ in~\eqref{eq:mu} is continuous, hence $C=\mu\inv((-\infty,\delta])$ and $D=\mu\inv([\delta,+\infty))$ are closed, because they are the inverse image of a closed set through a continuous function. The flow map $F$ is continuous and single-valued, thus, in particular, it is outer semicontinuous and locally bounded. Moreover, $F(x,y)$ is a singleton, therefore convex for each $(x,y)\in C$. Since $G(x,y)\ceq\{x\}\x\yminV(x)$, the first component of $G$ is the identity function. Therefore, $G$ is outer semicontinuous if and only if $\yminV$ is outer semicontinuous and it follows from Proposition~\ref{prop:muIsNice} that this is indeed the case. Moreover, since the identity is continuous and $\yminV(\sphere n)$ is compact, it follows that $G$ is locally bounded.
{It follows from Proposition~\ref{prop:muIsNice} that\linebreak\cite[Assumption~6.5]{goebel_hybrid_2012} is satisfied.}
\begin{comment}
From~\eqref{eq:closed_loop}, we conclude that
%
\begin{align*}
\inner{\grad V^{y}(x)}{F(x,y)}=-\norm{\PTSn(x)\grad V^{y}(x)}^2
\end{align*} 
for each $(x,y)\in C$. It follows from~\eqref{eq:controller} that, for each $(x,y)\in D$ and $g\in G(x,y)$, 
\begin{align*}
V(g)-V(x,y)&=V(x,\yminV(x))-V(x,y)\\
&=\minVy(x)-V(x,y)\\
&=-\gap(x,y)\leq-\delta.
\end{align*}
\end{comment}
{From computations similar to~\cite{mayhew_global_2013},} we conclude that the growth of $V$ along solutions to~\eqref{eq:closed_loop} is bounded by $u_C,u_D$, where
\begin{subequations}
\label{eq:uCuD}
\mathtoolsset{showonlyrefs=false}
\begin{align}
u_C(x,y)&\ceq\begin{cases}
-\norm{\PTSn(x)\grad{V^y(x)}}^2 & \text{ if } (x,y)\in C\\
-\infty & \text{ otherwise} 
\end{cases},\\
u_D(x,y)&\ceq\begin{cases}
-\delta & \text{ if } (x,y)\in D\\
-\infty & \text{ otherwise} 
\end{cases}
\end{align}
\end{subequations}
{for each $(x,y)\in\sphere n\x\yset$. It follows from the assumption that $V$ is a synergistic potential function relative to $r$ that $\gap(x,y)>\delta$ for each $(x,y)\in\sticky(V)$, hence $u_C(x,y)<0$ for each $(x,y)\in\sphere n\x\yset\minus\Amc$. Since $u_D(x,y)<0$ for all $(x,y)\in\sphere n\x\yset$, it follows from~\cite[Corollary~8.9]{goebel_hybrid_2012} that $\Amc$ is globally pre-asymptotically stable for~\eqref{eq:closed_loop}.
%
%It follows from continuity of $F$ that, for each $\xi\in C\minus D$, there exists $\epsilon> 0$ and a continuously differentiable function $z:[0,\epsilon]\to\sphere n\x\yset$ such that $z(0)=\xi$, $\dot z(t)=F(z(t))$ for all $t\in[0,\epsilon]$ and $z(t)\in C$ for all $t\in[0,\epsilon]$. 
Since $G(D)\subset C\cup D=\sphere n\x\yset$, $\sphere n\x\yset$ is compact and, for each $(x,y)\in C\minus D$,  $F(x,y)$ belongs to the tangent cone to $C$ at $(x,y)$, it follows from~\cite[Proposition~6.10]{goebel_hybrid_2012} that each maximal solution to~\eqref{eq:closed_loop} is complete.}
\qed\end{pf}
We are also able to show that, under some additional conditions, the set~\eqref{eq:Amc} is globally exponentially stable in the $t$-direction for the hybrid system~\eqref{eq:closed_loop}.

\begin{theorem}\label{theorem:expStability}
Given $r\in\sphere n$ and a compact set $\yset$, if there exists $V\in\spf[r]{\yset}$ with synergy gap exceeding $\delta>0$ satisfying the following conditions 
\begin{subequations}
\mathtoolsset{showonlyrefs=false}
\label{eq:expConditions}
\begin{align}
\label{eq:Vbounds}&\lbar{\alpha}\norm{x-r}^p\leq V(x,y)\leq \ubar{\alpha}\norm{x-r}^p & &\forall (x,y)\in C\cup D,\\
\label{eq:Vgamma}&\inner{\grad{V}(x,y)}{F(x,y)}\leq-\lambda V(x,y) & &\forall (x,y)\in C,
\end{align}
\end{subequations}
for some $p,\lbar{\alpha},\ubar{\alpha},\lambda>0$, then the set~\eqref{eq:Amc} is globally exponentially stable in the $t$-direction for the hybrid system~\eqref{eq:closed_loop}.
\end{theorem}
\begin{pf}
{It follows from the proof of Theorem~\ref{theorem:gas} that~\eqref{eq:closed_loop} satisfies~\cite[Assumption~6.5]{goebel_hybrid_2012}, each of its maximal solutions is complete and $V$ is nonincreasing during jumps. Since $\gap(g)=0$ for each $g\in G(D)$, we have that $G(D)\cap D=\emptyset$. It follows from~\eqref{eq:expConditions} that the conditions of~\cite[Theorem~1]{casau_hybrid_2017} are satisfied, thus $\Amc$ is globally exponentially stable in the $t$-direction for the hybrid system~\eqref{eq:closed_loop}.}
%Since $\delta>0$, it follows from~\eqref{eq:Vjump} that there is a strict decrease of $V$ during jumps.  
%The relation $G(D)\cap D=\emptyset$ follows from the fact that $\gap(g)=0<\delta$ for each $g\in G(D)$, thus it follows from~\cite[Lemma~2.7]{sanfelice_invariance_2007} that solutions to~\eqref{eq:closed_loop} are complete and non-Zeno. The desired result follows from Theorem~\cite[Theorem~1]{casau_hybrid_2017}.
%Conditions~\eqref{eq:Vbounds} and~\eqref{eq:Vgamma} match conditions~\eqref{eq:expV} and~\eqref{eq:expVdot}, respectively, hence the desired result follows from Theorem~\ref{thm:exp}.
\qed\end{pf}

\subsection{Construction of centrally synergistic potential function on $\sphere n$}\label{sec:construct}

Given $r\in\sphere n$, we construct a centrally synergistic potential function on $\sphere n$ using the height function:
%\begin{equation}\label{eqn:height}
%\begin{aligned}
$\height{r}(x)\ceq 1-r\tp x$ for all $x\in\sphere n.$
%\end{aligned}
%\end{equation}
Let $k > 0$, $\Vdom \ceq \Snsqr \minus \{(r,r)\}$, and define $V: \Vdom \to \R{}$ for all $(x,y) \in \Vdom$ as
\begin{equation}\label{eqn:V}
\begin{split}
V(x,y) &= \frac{\height{r}(x)}{h_{r}(x) + k\height{y}(x)} = \frac{1-r\tp  x}{1-r\tp  x + k\left(1-y\tp  x\right)}.
\end{split}
\end{equation}
%which is endowed with the following properties.

We now provide some differential properties of $V$, which follow from elementary calculation and some tedious manipulation, so they are presented without proof.
\begin{lemma}\label{lem:gradV}
The function $V: \Vdom \to [0,1]$ defined in \eqref{eqn:V} satisfies
\begin{subequations}
\mathtoolsset{showonlyrefs=false}
\begin{align}
\label{eqn:Vgradx}
\grad V^y(x) &= %\frac{\left(1-r\tp  x\right)y - \left(1-y\tp  x\right) r}
%{\left(1-r\tp  x + 1-y\tp  x\right)^{2}}\\
\frac{kV(x,y)y - (1-V(x,y))r}{1-r\tp  x + k\left(1-y\tp  x\right)}\\
\label{eqn:normVgradx}
\left| \PTSn(x)\grad V^y(x) \right|^{2} &= \frac{2kV(x,y)\left(1-V(x,y)\right)\left(1-r\tp  y\right)}
{\left(1-r\tp  x + k\left(1-y\tp  x\right)\right)^{2}}
\end{align}
\end{subequations}
\end{lemma}
Given $\yrbnd \in \R{}$ satisfying $-1 <\yrbnd < 1,$
we define the set $\yset \subset \sphere n$ as
\begin{equation}\label{eqn:yset}
\yset = \left\{ y \in \sphere n: r\tp  y \leq \yrbnd\right\}.
\end{equation}
The boundary of $\yset$ is $\partial\yset = \left\{y \in \sphere n: r\tp  y = \yrbnd\right\}.$

\begin{lemma}\label{lem:yminV}
Given $r\in\sphere n$ and $\yrbnd\in[-1,1)$, let $\yset$ be given by~\eqref{eqn:yset}. The following holds for the function $V$ given in~\eqref{eqn:V}
\begin{subequations}
\mathtoolsset{showonlyrefs=false}
 \begin{align}
\label{eqn:yminV}
\yminV(x) &= \begin{cases}
\yset & \text{if }x=r\\
-x & \text{if } r\tp  x \geq -\yrbnd \\
\frac{\sigma\left(r\tp  x\right)\PTSn(x)r}{\left|\PTSn(x)r\right|} + \alpha\left(r\tp  x\right)x & \text{if } -1 < r\tp  x < -\yrbnd \\
\partial\yset & \text{if } r\tp  x = -1.
\end{cases}\\
\label{eqn:cyminV}
\minVy(x) &= \begin{cases}
\displaystyle\frac{1-r\tp  x}{1-r\tp  x+2k} & \text{if } r\tp  x \geq -\yrbnd \\
\displaystyle\frac{1-r\tp  x}{1-r\tp  x +k\left(1-\alpha\left(r\tp  x\right)\right)} & \text{if } r\tp  x < -\yrbnd 
%\\ \frac{2}{2+k(1+\yrbnd)} & \text{if } x = -r.
\end{cases}
 \end{align}
 \end{subequations} 
for each $x \in \sphere n$, where 
\IfFull{%
\begin{subequations}
\mathtoolsset{showonlyrefs=false}
\begin{align}
\label{eqn:alpha} \alpha(v) &= \yrbnd v - \sqrt{\left(1-v^{2}\right)\left(1-\yrbnd^{2}\right)},\\
\label{eqn:sigma}\sigma(v) &= \yrbnd\sqrt{1-v^{2}} + v\sqrt{1-\yrbnd^{2}}
\end{align}
\end{subequations}}{%
$\alpha(v) = \yrbnd v - \sqrt{\left(1-v^{2}\right)\left(1-\yrbnd^{2}\right)}$ and
$\sigma(v) = \yrbnd\sqrt{1-v^{2}} + v\sqrt{1-\yrbnd^{2}}$}
for each $v\in[-1,1]$,
and $\minVy$, $\yminV$ are defined in~\eqref{eq:minVy} and~\eqref{eq:yminV}, respectively.\hfill $\square$
\end{lemma}
\begin{pf}
Suppose that $x = r$. Then, according to \eqref{eqn:argminV} of Lemma~\ref{lem:V}, $V(x,y) = 0$ for all $y \in \yset$. Thus, any $y \in \yset$ attains the minimum of $y \mapsto V(r,y) = 0$. 
If $r\tp  x \geq -\yrbnd$, it follows that $-x \in \yset$. In this case, \eqref{eqn:argminyV} of Lemma~\ref{lem:V} yields that the unconstrained minimizer of $y \mapsto V(x,y)$, which is $-x$, is also the minimizer of $y \mapsto V(x,y)$ when $y$ is constrained to $y \in \yset$. 
When $x = -r$, we have that
$V(-r,y) = 2/(2 + k(1+r\tp  y)).$
Clearly, $y \mapsto V(-r,y)$ is minimized by maximizing $r\tp  y$. When $y$ is constrained to $\yset$, the maximum value of $r\tp  y$ is $\yrbnd$, which is attained by any $y$ satisfying $r\tp  y = \yrbnd$, or equivalently, $y \in \partial \yset$.
We now examine the case when $-1 < r\tp  x \leq -\yrbnd$. Since $-x \notin \yset$, it suffices to study the solutions to the constrained minimization problem
\IfFull{\begin{equation}\label{eqn:Vminyprob}
\begin{aligned}[t]
\underset{y}{\text{minimize}} &\quad V(x,y) \\
\text{subject to} &\quad 1-y\tp  y = 0 \\ 
&\quad r\tp  y - \yrbnd = 0.
\end{aligned}
\begin{multlined}
\end{multlined}
\end{equation}}{$\min\left\{y\tp x:1-y\tp  y = 0,r\tp  y - \yrbnd = 0\right\}$}
by means of Lagrange multipliers.
\IfFull{%
Note that the constraints of the optimization problem are equivalent to $y \in \partial\yset$.  
Recalling \eqref{eqn:minyV2}, note that solutions to \eqref{eqn:Vminyprob} are equivalently found as solutions to the constrained minimization problem
\begin{equation}\label{eqn:Vminyprob2}
\begin{aligned}[t]
\underset{y}{\text{minimize}} &\quad y\tp  x \\
\text{subject to} &\quad 1-y\tp  y = 0 \\ 
&\quad r\tp  y - \yrbnd = 0.
\end{aligned}
\end{equation}
We proceed by constructing the Lagrangian, $L: \R{n+1} \times \R{} \times \R{} \to \R{}$ for each $(y,\lambda,\beta) \in \R{n+1} \times \R{} \times \R{}$ as
\begin{equation}\label{eqn:lagrangian}
L(y,\lambda,\beta) = y\tp  x + \lambda(y\tp  r - \yrbnd) + \frac{1}{2}\beta(1-y\tp  y).
\end{equation}
Setting $\grad L(y,\lambda,\beta) = 0$ yields
\begin{equation}\label{eqn:stationarity}
\beta y = x + \lambda r,
\end{equation}
in addition to the constraint that $y \in \yset$.
Seeing that $\beta = 0$ implies that $x \in \{r,-r\}$ (where we have already solved the problem at hand), we henceforth assume that $\beta \neq 0$ and look for constants $a,b \in \R{}$ such that
\begin{equation}\label{eqn:solvefory}
y = ax + b r.
\end{equation}
To ease notation, we let $v = r\tp  x$ for the remainder of the proof. Taking the norm of both sides of \eqref{eqn:solvefory} and then noting that $|y| = |x| = |r| = 1$, yields
\begin{equation}
\label{eqn:ynorm}
1 = a^{2} + 2abv + b^{2}.
\end{equation}
Then, by multiplying \eqref{eqn:solvefory} on the left by $r\tp  $ and applying the constraint that $r\tp  y = \yrbnd$, we arrive at
\begin{equation}
\label{eqn:ry}
\yrbnd = av + b.
\end{equation}
Solving \eqref{eqn:ry} for $b$ and substituting the result into \eqref{eqn:ynorm} yields (with some rearrangement that
\begin{equation}
a^{2} = \frac{1-\yrbnd^{2}}{1-v^{2}}.
\end{equation}
By multiplying \eqref{eqn:solvefory} on the left by $x\tp  $, we see that
\begin{equation}
x\tp  y = a\left(1-v^{2}\right)+\yrbnd v.
\end{equation}
Since $V(x,y)$ is minimized when $x\tp  y$ is minimized and $1-v^{2} = 1-(r\tp  x)^2 \geq 0$, it follows that $V$ is minimized when
\begin{equation}
a = -\sqrt{\frac{1-\yrbnd^{2}}{1-v^{2}}}.
\end{equation}
By \eqref{eqn:ry},
\begin{equation}\label{eqn:bsolved}
b = \yrbnd + v\sqrt{\frac{1-\yrbnd^{2}}{1-v^{2}}}.
\end{equation}
Now, let 
\begin{align}
\label{eqn:rcoeff}
\begin{split}
\sigma &= b\sqrt{1-v^{2}}\\
&= \yrbnd\sqrt{1-v^{2}} + v\sqrt{1-\yrbnd^{2}}
\end{split} \\
\label{eqn:xcoeff}
\begin{split}
\alpha &= a + vb \\
&= v\yrbnd - \sqrt{\left(1-\yrbnd^{2}\right)\left(1-v^{2}\right)}
\end{split}
\end{align}
Note that $\norm{\PTSn(x)r} = \norm{\left(I-xx\tp  \right)r} = \sqrt{1-(r\tp  x)^{2}}$. Then, by some rearrangement, it follows that
\begin{equation}\label{eqn:ynew}
\begin{split}
y &= (\alpha - vb)x + b r \\
&= \alpha x + b(I-xx\tp  )r \\
&= \alpha x + \sigma \frac{\PTSn(x)r}{\norm{\PTSn(x)r}}r.
\end{split}
\end{equation}
Recalling the definitions of $\alpha$ and $\sigma$ in \eqref{eqn:alpha} and \eqref{eqn:sigma}, respectively, yields the desired result.%
}{}%endif
\qed\end{pf}

\IfFull{%
Moreover, from Corollary~\ref{cor:critV} and from definition~\eqref{eq:stickyV} it follows
\begin{equation}
\label{eqn:sticky}
\sticky(V) = \left\{(x,x)\in\sphere n\x\sphere n: x \in \yset \right\},
\end{equation}
for $V$ given by~\eqref{eqn:V}. In the sequel, let 
\begin{equation}
\label{eqn:ss}
\begin{aligned}
\Xmc:=\sphere n\x\yset,
\end{aligned}
\end{equation}
denote the reduced state space and let the function $\stickygap : \sphere n \to \R{}$ be given by
\begin{equation}\label{eqn:stickygap}
\begin{split}
\stickygap(x) &:= 1-\minVy(x).
\end{split}
\end{equation}
for all $x \in \sphere n$. We now note that both $\gap$ and $\stickygap$ are continuous on $\Xmc$ and $\sphere n$, respectively, and that $\stickygap$ agrees with $\gap$ on the set $\sticky$ (when the duplicated argument is ignored). 
\begin{theorem}\label{thm:gap}
The functions $\gap : \Xmc \to \R{}$ and $\stickygap : \sphere n \to \R{}$ are continuous and for each $x \in \yset$ (and therefore each $(x,x) \in \sticky(V)$),
\begin{equation}\label{eqn:stickygapagree}
\gap(x,x) = \stickygap(x).
\end{equation}
Furthermore, $\stickygap(x) > 0$ for all $x \in \sphere n$, and in particular,
\begin{equation}\label{eqn:minstickygap}
\min_{x \in \yset} \stickygap(x) = \min_{x \in \sphere n} \stickygap(x) = \frac{1+\yrbnd}{2/k+1+\yrbnd}
\end{equation}
\hfill $\square$
\end{theorem}
\begin{pf}
The continuity of $\gap$ and $\stickygap$ follow from the continuity of $\minVy$, which follows from the explicit calculation in \eqref{eqn:cyminV} of Lemma~\ref{lem:yminV}. 
Recalling that $V(x,x) = 1$ for all $(x,x) \in \Vdom$ from \eqref{eqn:argmaxV} of Lemma~\ref{lem:V} and the calculation of $\minVy$ in \eqref{eqn:cyminV}, we calculate $\gap(x,x)$ for each $x \in \yset$ as
\begin{equation}\label{eqn:stickygap2}
\begin{split}
\gap(x,x) &= V(x,x) - \minVy(x) \\
&= 1-\minVy(x).
\end{split}
%\begin{cases}
%1-\frac{1-r\tp  x}{1-r\tp  x+2k} & -\yrbnd \leq r\tp  x \leq \yrbnd \\
%1-\frac{1-r\tp  x}{1-r\tp  x + k\left(1-\alpha\left(r\tp  x\right)\right)} & -1 \leq r\tp  x < -|\yrbnd|.
%\end{cases}
\end{equation}
We now show that $\stickygap(x) > 0$ for all $x \in \sphere n$ by finding an explicit lower bound. {Let}
\begin{align}
\label{eqn:gapterms1}
\eta_{1}(v) &= \frac{1-v}{1-v+2k} \\
\label{eqn:gapterms2}
\eta_{2}(v) &= \frac{1-v}{1-v+k\left(1-\alpha\left(v\right)\right)} \\
\label{eqn:gapterms3}
\eta_{3} &= \frac{2}{2+k(1+\yrbnd)}.
\end{align}
It follows {from~\eqref{eqn:yminV} and~\eqref{eqn:cyminV}} that
\begin{equation}\label{eqn:stickygapmin}
\min_{x\in\yset} \stickygap(x) = 1 - 
\max\left\{
\max_{|v|\leq \yrbnd} \eta_{1}(v),
\max_{-1 \leq v \leq -\yrbnd} \eta_{2}(v),
\eta_{3}
\right\}
\end{equation}
First, we calculate the derivative of $\eta_{1}$ as
\begin{equation}
\eta_{1}'(v) = \frac{-2k}{\left(1-v+2k\right)^{2}}.
\end{equation}
Since the denominator of $\eta_{1}'(v)$ above is positive for all $v \in [-1,1]$ and because $k >0$, $\eta_{2}'(v)$ is clearly negative for all $v \in [-1,1]$. Hence,
\begin{equation}\label{eqn:gapterm1min}
\max_{|v|\leq \yrbnd} \eta_{1}(v) = \eta_{1}(-\yrbnd) = \frac{1+\yrbnd}{1+\yrbnd+2k}
\end{equation}
After some effort, the derivative of $\eta_{2}$ can be calculated as
\begin{equation}
\eta_{2}'(v) = k\frac{(1-v)\alpha'(v)-1+\alpha(v)}
{\left(1-v+k\left(1-\alpha\left(v\right)\right)\right)^{2}}.
\end{equation}
Since the denominator of $\eta_{2}'(v)$ above is always positive for $v \in [-1,1]$ and $k > 0$, the sign of $\eta_{2}'(v)$ can be determined from the sign of the numerator. In this direction, we have
\begin{equation}
\begin{multlined}
(1-v)\alpha'(v)-1+\alpha(v) \\
= \yrbnd-1-\sqrt{(1-v^{2})(1-\yrbnd^{2})} - v(1-v)\frac{\sqrt{1-\yrbnd^{2}}}{\sqrt{1-v^{2}}}.
\end{multlined}
\end{equation}
We note the equivalent series of inequalities (for all $v \in (-1,1)$),
\begin{equation}
\begin{aligned}
&\yrbnd-1-\sqrt{(1-v^{2})(1-\yrbnd^{2})} - v(1-v)\frac{\sqrt{1-\yrbnd^{2}}}{\sqrt{1-v^{2}}} <0 \\
&\begin{multlined}
(1-\yrbnd)\sqrt{1-v^{2}} + (1-v^{2})\sqrt{1-\yrbnd^{2}}\\
+v(1-v)\sqrt{1-\yrbnd^{2}} > 0 \end{multlined}\\
&(1-\yrbnd)\sqrt{1-v^{2}} + (1-v)\sqrt{1-\yrbnd^{2}} >0.
\end{aligned}
\end{equation}
Since $|\yrbnd| < 1$ and $|v| < 1$, it follows that the inequality above is satisfied for all $v \in (-1,1)$, so $\eta_{2}'(v)$ is negative for all $v \in (-1,-\yrbnd)$. Noting that $\alpha(-\yrbnd) = -1$ and $\alpha(-1) = -\yrbnd$, we have
\begin{equation}
\min_{-1 \leq v \leq -\yrbnd} \eta_{2}(v) = \eta_{2}(-\yrbnd) = \eta_{1}(-\yrbnd) = \frac{1+\yrbnd}{1+\yrbnd+2k}
\end{equation}
and
\begin{equation}
\max_{-1 \leq v \leq -\yrbnd} \eta_{2}(v) = \eta_{2}(-1) = \eta_{3} = \frac{2}{2+k(1+\yrbnd)}.
\end{equation}
Thus,
\begin{equation}
\min_{x \in \yset} \stickygap(x) = 1-\frac{2}{2+k(1+\yrbnd)} = \frac{1+\yrbnd}{2/k+1+\yrbnd}.
\end{equation}
\qed\end{pf}
}{} %endif

Using~\eqref{eqn:cyminV}, we may compute~\eqref{eq:mu}, from which the next result follows.

\begin{corollary}
 For any given $r\in\sphere n$ and {$\yrbnd\in (-1,1)$,} the function~\eqref{eqn:V} is a centrally synergistic potential function relative to $r$ with synergy gap exceeding $\delta$, for any $\delta\in \left(0,(1+\yrbnd)/(2/k+1+\yrbnd)\right).$
\end{corollary}
\begin{pf}
This result follows from the fact that
\IfFull{%
\begin{equation}\label{eq:minmu}
\begin{aligned}
\min\{\gap(x,y)&:(x,y)\in\crit V^y\minus\{r\}\x\yset\}\\
&=\min\{\gap(x,x):x\in\yset\}\\
&=\min\{1-\minVy(x):x\in\yset\}\\
&=\frac{1+\yrbnd}{2/k+1+\yrbnd}.
\end{aligned}
\end{equation}}{%
\begin{equation}\label{eq:minmu}
\begin{aligned}
\min\{\gap(x,y)&:(x,y)\in\crit V^y\minus\{r\}\x\yset\}\\
%&=\min\{\gap(x,x):x\in\yset\}\\
%&=\min\{1-\minVy(x):x\in\yset\}\\
&=\frac{1+\yrbnd}{2/k+1+\yrbnd}.
\end{aligned}
\end{equation}}
\end{pf}

In addition to global asymptotic stability of $\Amc$ for~\eqref{eq:closed_loop},  we also show below that the function~\eqref{eqn:V} satisfies~\eqref{eq:expConditions}, thus it follows from Theorem~\ref{theorem:expStability} that global $\Amc$ is also globally exponentially stable in the $t$-direction for~\eqref{eq:closed_loop}. It follows from the fact that $\sticky(V)=V\inv (1)$ (c.f. Lemma~\ref{lem:V}) and from $\sticky(V)\subset D$, with $D$ given in~\eqref{eq:controller}, that $\maxVflow\in\R{}$, given by $\maxVflow = \max_{(x,y) \in C} V(x,y),$
with $C$ given in~\eqref{eq:controller} satisfies $0 < \maxVflow < 1$. We note that $\maxVflow$ exists since $V$ is continuous on the compact set $C$.
The next theorem follows naturally from these considerations

\begin{theorem}\label{thm:VexpConditions}
The function $V\in\C^1(\sphere n\x\yset)$ given in~\eqref{eqn:V} satisfies~\eqref{eq:expConditions} with
\begin{subequations}
\mathtoolsset{showonlyrefs=false}
\label{eq:aagg}
\begin{align}
\label{eq:laa}\lbar{\alpha}&=(2(1+k+\sqrt{1+2k\yrbnd+k^2}))^{-1},\\
\label{eq:uaa}\ubar{\alpha}&=(2(1+k-\sqrt{1+2k\yrbnd+k^2}))^{-1}\\
\label{eq:gg}\lambda&=\frac{2k(1-\maxVflow)(1-\yrbnd)}{\left(1+k+\sqrt{1+2k\yrbnd+k^2}\right)^{2}}.
\end{align}
\end{subequations}
Therefore, the set~\eqref{eq:Amc} is globally exponentially stable in the $t$-direction for~\eqref{eq:closed_loop}.
\end{theorem}
\begin{pf}
Since $h_r(x)=1-r\tp   x=\norm{x-r}^2/2$ for every $x\in\sphere n$, it follows from~\eqref{eqn:globaldenombnd} that~\eqref{eqn:V} satisfies~\eqref{eq:Vbounds} with $\lbar{\alpha}$ and $\ubar{\alpha}$ given by~\eqref{eq:laa} and~\eqref{eq:uaa}, respectively. Since $r\tp   y\leq\yrbnd$ for each $(x,y)\in\sphere n\x\yset$, then $1-r\tp   y\geq 1-\yrbnd$. It follows from the fact that $V(x,y)\leq \maxVflow$ for each $(x,y)\in C$ and from Lemma~\ref{lem:globaldenombnd} that
\begin{equation}\label{eqn:gradVboundflow}
\left| \PTSn(x)\grad V^y (x) \right|^{2} \geq \frac{2k(1-\maxVflow)(1-\yrbnd)}{\left(1+k+\sqrt{1+2k\yrbnd+k^{2}}\right)^{2}}V(x,y),
\end{equation}
which proves that~\eqref{eq:Vgamma} is satisfied. Global exponential stability of~\eqref{eq:Amc} for~\eqref{eq:closed_loop} follows from Theorem~\ref{theorem:expStability}.
\qed\end{pf}
\subsection{Minimal Geodesics}
We have shown that the proposed synergistic potential function~\eqref{eqn:V}
ensures not only global asymptotic stability of $\Amc$ for~\eqref{eq:closed_loop}, but also global exponential stability. Moreover, there is a third property of this function that is worth mentioning: for each $x_0\in D$, the flows generated by the gradient vector field $\PTSn(x)\grad V^{\yminV(x_0)}(x)$ converge to $r$ through the minimal geodesic, i.e., the path of minimum distance between $x_0$ and $r$, as proved next.
For every pair of points $p,q\in M$, where $M$ is a compact manifold, there is $\geo :\R{}\rightarrow M$ with $\geo(a)=p$ and $\geo(b)=q$ for some $a,b\in R{}$ such that the length of $\geo$, given by $\length{\geo}=\int_a^b\norm{\frac{dc}{dt}(\tau)}d\tau,$
is the lowest among all other paths with endpoints $p$ and $q$ (c.f.~\cite{burns_differential_2005}). If a path $p:[0,\infty)\rightarrow M$ is Lebesgue integrable, then its length is given by $\length[\infty]{p}=\int_0^\infty\norm{\frac{dp}{dt}(\tau)}d\tau.$
Considering the Riemannian manifold $\sphere n$ with the standard Euclidean metric, its (unit-velocity) geodesics are of the form $x(t)=a\cos t+b\sin t,$
for each $t\in\R{}$, with $a,b\in\sphere n$ satisfying $\inner{a}{b}=0$, as shown in~\protect{\cite[Example~30]{petersen_riemannian_2006}}. Given $r\in\sphere n$, one verifies that the minimal geodesic between $r$ and any given point $x_0\in\sphere n\backslash\{r,-r\}$ is given by
%
%\begin{equation}
%\label{eq:segment}
%\begin{aligned}
$\geo (t)=x_0\cos t+\PTSn(x_0)r\sin t/\norm{\PTSn(x_0)r},$
%\end{aligned}
%\end{equation}
%
for each \protect{$t\in [0,\arccos(x_0\tp   r)]$}. If $x_0=-r$, then each minimal geodesic from $x_0$ to $r$ is given by $\geo(t)=x_0\cos t+x^\bot\sin t,$ 
for each $t\in[0,\pi]$, where $x^\bot\in\{y\in\sphere n: \inner{y}{x_0}=0\}$. {In particular, if $x_0$ is antipodal to $r$ and $n>1$ then there are uncountably many minimal geodesics from $x_0$ to $r$. }

 It is possible to verify that, for every solution $(x,y)$ of~\eqref{eq:closed_loop} and each $(t,j),(t,j+1)\in\dom{(x,y)}$, the following holds: $x(t,j)=x(t,j+1).$ Therefore, $x{\downarrow_t} (t):=x(t,J(t)),$ is defined for each $t\in [0,\sup_t\dom{(x,y)})$, for each solution $(x,y)$ to~\eqref{eq:closed_loop}, where with $J(t):=\max\{j:(t,j)\in\dom (x,y)\}.$ Moreover, it is absolutely continuous, hence $\length[\infty]{x{\downarrow_t}}$ is well-defined. Choosing $V$ in~\eqref{eqn:V}, we show next that the solutions to the hybrid system~\eqref{eq:closed_loop} that start in the jump set have minimal length.

\begin{lemma}\label{lem:geo}
Consider the function $V\in\C^1(\sphere n\x\yset)$ given in~\eqref{eqn:V}. For each solution $(x,y)$ to~\eqref{eq:closed_loop} with initial condition $(x_0,y_0)\in D$, we have that $\length[\infty]{x_t}=\length{\geo_{x_0,r}}$, where $\geo_{x_0,r}$ denotes the minimal geodesic from $x_0$ to $r$.
\end{lemma}
\begin{pf}
This result follows from the fact that the vector field $\PTSn(x)\grad V^{\yminV(x_0)}(x)$ is tangent to the geodesic connecting $r$ and $x_0$ and from Theorem~\ref{thm:VexpConditions} which shows that solutions converge to $r$.
\IfFull{%
Firstly, notice that, for each $a,b\in\sphere n$ satisfying $\inner{a}{b}=0$, we have that the charts $(\geo_\pm\inv,O_\pm)$, given by
\begin{equation}
\label{eq:S1}
\begin{aligned}
\geo_+&:=a\cos t+b\sin t & O_+:=c_+(\I),\\
\geo_-&:=-a\cos t+b\sin t & O_-:=c_-(\I),
\end{aligned}
\end{equation}
where $\I:=(-\pi,\pi)$, define an one-dimensional submanifold of $\sphere n$, which we denote by $\sphere 1(a,b)$. Moreover, for each $r\in\sphere n$ and for each $x_0\in \sphere n\backslash\{r,-r\}$, the maximal geodesic through $x_0$ and $r$, given by
\begin{equation}
\label{eq:maximal}
\begin{aligned}
\geo (t)=x_0\cos t+\frac{\PTSn(x_0)r}{\norm{\PTSn(x_0)r}}\sin t,
\end{aligned}
\end{equation}
 and defined for each $t\in\R{}$, belongs to $\sphere 1(x_0,\frac{\PTSn(x_0)r}{\norm{\PTSn(x_0)r}})$, and the map~\eqref{eq:maximal} is a smooth immersion because its push-forward does not vanish for any $t\in\R{}$. Since the initial condition of the solution $(x,y)$ belongs to the jump set, we have that $(0,1)\in\dom{(x,y)}$, and $(x,y)(0,1)=(x_0,\yminV(x_0))$. The desired result follows from: the vector field $\PTSn(x)\grad V^{\yminV(x_0)}(x)$ is tangent to $\sphere 1(x_0,\frac{\PTSn(x_0)r}{\norm{\PTSn(x_0)r}})$; from Theorem~\ref{thm:VexpConditions} we have that solutions converge to $r$.
The former condition is verified if $\grad V^{\yminV(x_0)}(x)$ can be written as a linear combination of $x_0$ and $\frac{\PTSn(x_0)r}{\norm{\PTSn(x_0)r}}$ for each $x\in\sphere 1(x_0,\frac{\PTSn(x_0)r}{\norm{\PTSn(x_0)r}})$, because, in that case $\PTSn(x)\grad V^{\yminV(x_0)}(x)\in\T_x\sphere 1(x_0,\frac{\PTSn(x_0)r}{\norm{\PTSn(x_0)r}})$. 
It is possible to rewrite~\eqref{eqn:Vgradx} as follows $\grad V^y(x)= k\frac{y(1-r\tp   x)-r(1-x\tp   y)}{(1-r\tp   x+k(1-y\tp   x))^2},$
thus $\grad V^y(x)$ is a linear combination of $y$ and $r$. Notice that $r$ can be written as a linear combination of $x_0$ and $\frac{\PTSn(x_0)r}{\norm{\PTSn(x_0)r}}$ because $r = \PTSn(x_0)r+x_0x_0\tp r$. On the other hand, from the definition of $\yminV$ given in~\eqref{eqn:yminV}, we have that
%\begin{comment}%
\begin{equation}
\begin{aligned}
\yminV(x_0) &= \begin{cases}
\yset & \text{if }x_0=r\\
-x_0 & \text{if } r\tp  x_0 \geq -\yrbnd \\
\rho(x_0) & \text{if } -1 < r\tp  x < -\yrbnd \\
\partial\yset & \text{if } r\tp  x = -1.
\end{cases}
\end{aligned}
\end{equation}
with
$\yminV(x_0)\ceq\sigma\left(r\tp  x_0\right)\frac{\PTSn(x_0)r}{\left|\PTSn(x_0)r\right|} + \alpha\left(r\tp  x_0\right)x_0$ for each $x_0\neq \pm r$ 
%\end{comment}
$\yminV(x_0)$ is also a linear combination of $x_0$ and  $\frac{\PTSn(x_0)r}{\norm{\PTSn(x_0)r}}$ for any $x_0$ satisfying $-1<r\tp x_0$ because $\alpha$ and $\sigma$ are scalar maps. Also, $r\tp x_0\neq 1$ because in that case $x_0=r\not\in D$. If $r\tp x_0=-1$, then any choice $y\in\partial\yset$ satisfies the required condition for some geodesic.%
}{}
\qed\end{pf}
Note that function $\height{r}$ induces a gradient-vector field that generates flows along geodesics for almost all initial conditions. However, unlike the aforementioned continuous feedback law, the controller presented in this section exponentially renders a reference point globally asymptotically stable, which is not possible with continuous feedback. Moreover, if we resolve the ambiguity at $x=-r$ by means of some discontinuity, we still are left without any guarantees of robustness to small measurement noise. Since the hybrid controller presented in this section satisfies the hybrid basic conditions~\cite[Assumption~6.5]{goebel_hybrid_2012}, the property of global asymptotic stability of $\Amc$ for~\eqref{eq:closed_loop} is endowed with robustness to small measurement noise, as discussed in~\cite{goebel_hybrid_2012}.

\section{Application to Trajectory Tracking for a Quadrotor}\label{sec:quad}
\IfHub{}{In this section, we apply the controller proposed in Section~\ref{sec:synergistic} to the problem of trajectory tracking for a quadrotor vehicle, i.e., an aerial vehicle with four counter rotating rotors that are aligned with a direction which is fixed relative to the body of the vehicle, as described in~\cite{hamel_dynamic_2002}.} The dynamics of a thrust vectored vehicle such as a quadrotor can be described by
\begin{subequations}
\label{eq:quad}
\mathtoolsset{showonlyrefs=false}
\begin{align}
\dot p&=v\\
\dot v&=\Rmat r u+\grv\\
\label{eq:dotR}\dot \Rmat&=\Rmat\sk(\omega)
\end{align}
\end{subequations}
where $p\in\R 3$ and $v\in\R 3$ denote the position and the velocity of the vehicle with respect to the inertial reference frame (in inertial coordinates), $\Rmat\in\SO(3)\ceq\{\Rmat\in\R{3\x 3}:\Rmat\tp \Rmat =\eye{3},\det(\Rmat)=1\}$ is the rotation matrix that maps vectors in body-fixed coordinates to inertial coordinates, $g\in\R{3}$ represents the gravity vector and $r\in\sphere 2\ceq\{x\in\R 3:\norm{x}=1\}$ is the thrust vector in body-fixed coordinates. Furthermore, the inputs to~\eqref{eq:quad} are $\wb\in\R 3$ and $u\in\R{}$ which represent the angular velocity in body-fixed coordinates and the magnitude of the thrust, respectively. The dynamical model~\eqref{eq:quad} is a simplification of the one provided in~\cite{hamel_dynamic_2002} that better suits our experimental setup, since there the \emph{Blade 200 QX} quadrotor that is used in the experiments has an embedded controller that tracks angular velocity and thrust commands. %and $$\sk(\omega)\ceq\bmtx{0 & -\omega_3 & \omega_2\\ \omega_3 & 0 & -\omega_1\\ -\omega_2 & \omega_1 & 0}$$ satisfying $\PTSn(\omega)=-\sk(\omega)^2$ for each $\omega=(\omega_1,\omega_2,\omega_3)\in\R 3$ (c.f.~\cite{hamel_dynamic_2002}). 
Furthermore, we assume that the reference trajectory satisfies the following assumption.

\begin{assumption}\label{ass:pd}
The reference trajectory $t\mapsto p_d(t)$ is defined for each $t\geq 0$ and there exist $M_2\in(0,\norm{\grv})$ and $M_3>0$ such that \IfHub{%
\begin{align*}
\norm{\ddot p_d(t)}\leq M_2\text{ and }p_d^{(3)}(t)\leq M_3
\end{align*}
for all $t\geq 0$.}{%
$\norm{\ddot p_d(t)}\leq M_2$ and $\norm{p_d^{(3)}(t)}\leq M_3$ for all $t\geq 0$.
}
\end{assumption}

Given a path that satisfies Assumption~\ref{ass:pd}, we define the tracking errors as \IfHub{
\begin{align*}
\pe&\ceq p-p_d\\
\ve&\ceq v-\dot p_d
\end{align*}}{$\pe\ceq p-p_d$ and $\ve\ceq v-\dot p_d,$}
whose dynamics can be derived from~\eqref{eq:quad} and are given by:
\IfHub{
\begin{equation}
\label{eq:errorsys}
\begin{aligned}
\dot\pe&=\ve, \\
\dot\ve&=\Rmat r u+\grv-\ddot p_d.
\end{aligned}
\end{equation}
}{%
\begin{equation}
\label{eq:errorsys}
\begin{aligned}
\dot\pe&=\ve, &
\dot\ve&=\Rmat r u+\grv-\ddot p_d.
\end{aligned}
\end{equation}
}%endif

Given $w:\R 6\to\R 3$, we define
\begin{subequations}
\label{eq:Tdu}
\mathtoolsset{showonlyrefs=false}
\begin{align}
\label{eq:Td}
\Td(z)&\ceq\frac{w(\pe,\ve)-\grv+\ddot p_d}{\norm{w(\pe,\ve)-\grv+\ddot p_d}}\\
\label{eq:ku}
\uctrl(z,\Rmat)&\ceq r\tp\Rmat\tp(w(\pe,\ve)-\grv+\ddot p_d)
\end{align}
\end{subequations}
for each $z\ceq(\ddot p_d,\pe,\ve)\in\trajset\x\R 6$ and $\Rmat\in\SO(3).$
Note that, if $\Rmat r=\Td(z)$ and $u=\uctrl(z,\Rmat)$, we obtain
\IfHub{%
\begin{equation}
 \label{eq:doubleint}
 \begin{aligned}
 \dot\pe&=\ve, \\
 \dot\ve&=w(\pe,\ve)
 \end{aligned}
 \end{equation} 
 }{%
\begin{equation}
 \label{eq:doubleint}
 \begin{aligned}
 \dot\pe&=\ve, &
 \dot\ve&=w(\pe,\ve)
 \end{aligned}
 \end{equation} 
}%endif
from~\eqref{eq:errorsys}, provided that 
\begin{equation}
\label{eq:nonzero}
\begin{aligned}
w(\pe,\ve)-\grv+\ddot p_d\neq 0.
\end{aligned}
\end{equation}
On the other hand, if $\Rmat r\neq \Td(z),$ then $\uctrl(z,\Rmat)$ is the solution to the least-squares problem: $$\min\{\norm{\Rmat r u+\grv-\ddot p_d-w(\pe,\ve)}^2:u\in\R{}\}.$$ The mismatch between $\Rmat r$ and $\Td(z)$ may lead to an increase in the the position and velocity tracking errors until the thrust vector $\Rmat r$ is aligned with $\Td(z)$.
The controller design is a two-step process, where we start by designing a feedback law $(\pe,\ve)\mapsto w(\pe,\ve)$ that exponentially stabilizes the origin of~\eqref{eq:doubleint} and then we design a partial attitude tracking controller that exponentially stabilizes $\Td(z)$ for the dynamics of the thrust vector $\Rmat r$. %This controller for the position subsystem is such that, if $\Rmat r_1 u= w(\pe,\ve)-\grv+\ddot p_d$ for each $(\ddot p_d,\pe,\ve)\in\R 9$ satisfying~\eqref{eq:nonzero}, then the thrust vector serves three purposes: applies feedback of the position and velocity errors, compensates gravity and applies a feedforward action of the desired acceleration. We assume that the thrust magnitude is an input to which we assign the feedback law~\eqref{eq:ku}, but we do not consider $\Rmat r_1$ as an input, thus we design an attitude controller that is able to track a vector reference $z\mapsto \Td(z)$ using the controller proposed in Section~\ref{sec:synergistic}. %A secondary control objective is attained by tracking a reference $\yaw\in\sphere 2$ that is orthogonal to $\Td$. 

%In the sequel, we develop a position tracking controller and two attitude controllers: one that tracks the commanded thrust vector and another that also tracks a yaw reference.

\subsection{Controller for the Position Subsystem}

We show next that it is possible to satisfy~\eqref{eq:nonzero} and exponentially stabilize the origin of~\eqref{eq:doubleint} from an arbitrary compact set $U$ using a continuously differentiable saturated linear feedback law 
\begin{equation}
\label{eq:w}
\begin{aligned}
w(\pe,\ve)\ceq \Sat_\bound\left(K\bmtx{\pe\tp & \ve\tp}\tp\right)
\end{aligned}
\end{equation}
satisfying $\Sat_\bound(x)=x$ for all $\norm{x}\leq \bound$ and $\norm{\Sat_\bound(x)}< \norm{\grv}-M_2$ for each $x\in\R{3}$ where $b\in(0,\norm{\grv}-M_2)$. %and $$v\mapsto\Sat_\bound(v)\ceq\bmtx{\sat_\bound(v_1) & \sat_\bound(v_2) & \sat_\bound(v_3)}\tp$$ for each $v\in\R 3$ with $\sat_\bound:\R{}\to\R{}$ continuously differentiable and nondecreasing, $. 
Defining $\sublevel{P}(\ell)\ceq \{x\in\R n:x\tp P x\leq \ell\}$ for $P\in\R{n\x n}$ and $\ell>0$, if we select $(H,\ell_H)\in\defpos{6}\x\Rpos$ such that $\sublevel{H}(\ell_H)$ is a bounding ellipsoid for $U$, then we guarantee that the {bounds $\pm b$ of the saturation function~\eqref{eq:w} are not reached} if there exists $(P,\ell_P)\in\defpos{6}\x \Rpos$ and $K\in\R{3\x 6}$, such that $\sublevel{P}(\ell_P)$ is forward invariant for every solution to~\eqref{eq:doubleint} from $U$ and
{\begin{equation}
\label{eq:Hellip}
\begin{aligned}
U\subset\sublevel{H}(\ell_H)\subset \sublevel{P}(\ell_P)\subset\sublevel{K\tp K}(\bound^2).
\end{aligned}
\end{equation}}
For each compact set $U\subset\R{6}$, it is possible to select controller parameters that satisfy~\eqref{eq:Hellip} as well as 
\begin{equation}
\label{eq:QR}
\begin{aligned}
(A+BK)\tp P+P(A+BK)\vleq -\Qlqr-K\tp \Rlqr K\\
\end{aligned}
\end{equation}
where 
\begin{equation}
\label{eq:AB}
\begin{aligned}
A&\ceq \bmtx{0 & \eye{3}\\ 0 & 0}, & B&\ceq \bmtx{0\\ \eye{3}}.
\end{aligned}
\end{equation}
thus guaranteeing also the exponential stability of the origin of~\eqref{eq:doubleint} from $U$, as stated in the proposition below.

\begin{proposition}\label{pro:ineqs}
{For each compact set $U\subset\R 6$}, $\bound>0$, $\ell_P>0$, $\ell_H>0$,  $\Rlqr\in\defpos{3}$ and $H\in\defpos{6}$, there exists $K\in\R{3\x 6}$ and $\Qlqr,P\in\defpos{6}$ such that the conditions~\eqref{eq:Hellip} and~\eqref{eq:QR} are satisfied.
\end{proposition}
\begin{pf}
From Lemma~4.20 in~\cite{saberi_internal_2012}, it follows that {there is a function} $\epsilon\mapsto \Qlqr(\epsilon)$, with the properties
\begin{equation}
\label{eq:Qeps}
\begin{aligned}
\Qlqr(\epsilon)&\vg 0 &\text{and } \frac{d \Qlqr(\epsilon)}{d \epsilon}&\vg 0
\end{aligned}
\end{equation}
for each $\epsilon\in(0,1]$, {that} generates a unique solution $P(\epsilon)\vg 0$ to
\begin{equation}
\label{eq:APPAeps}
\begin{aligned}
 A\tp P(\epsilon)+P(\epsilon) A-P(\epsilon) B\Rlqr\inv B\tp P(\epsilon) + \Qlqr(\epsilon)=0
\end{aligned}
\end{equation}
satisfying $P(\epsilon)\to 0$ as $\epsilon\to 0$. 

{Choosing $K=-\Rlqr\inv B\tp P(\epsilon) $, it follows that
\begin{equation}
\label{eq:APPA}
\begin{multlined}
(A+BK)\tp P(\epsilon)+P(\epsilon)(A+BK) \\
= A\tp P(\epsilon)+P(\epsilon) A-2P(\epsilon) B\Rlqr\inv B\tp P(\epsilon).
\end{multlined}
\end{equation}
Adding and subtracting $\Qlqr(\epsilon)$ to the right hand side of~\eqref{eq:APPA} and using~\eqref{eq:APPAeps} yields}
\begin{equation}
\label{eq:riccati}
\begin{multlined}
A\tp P(\epsilon)+P(\epsilon) A +K\tp B\tp P(\epsilon)+P(\epsilon) BK=\\
-P(\epsilon) B \Rlqr\inv B\tp P(\epsilon)-\Qlqr(\epsilon).
\end{multlined}
\end{equation}
Therefore, the condition~\eqref{eq:QR} is satisfied {with $P=P(\epsilon)$ and $\Qlqr=\Qlqr(\epsilon)$}. The condition $\sublevel{H}(\ell_H)\subset\sublevel{P}(\ell_P)$ is satisfied if and only if
\begin{equation}
\label{eq:PH}
\begin{aligned}
\frac{P(\epsilon)}{\ell_P}\vleq \frac{H}{\ell_H},
\end{aligned}
\end{equation}
and the condition $\sublevel{P}(\ell_P)\subset\sublevel{K\tp K}(\bound^2)$ is satisfied if and only if $\frac{K\tp K}{\bound^2}\vleq \frac{P(\epsilon)}{\ell_P}\iff P(\epsilon)^{\frac{1}{2}} B \Rlqr^{-2}B\tp P(\epsilon)^{\frac{1}{2}}\vleq \frac{\bound^2}{\ell_P}\eye{6},$ which is equivalent to
\begin{equation}
\label{eq:svmaxPeps}
\begin{aligned}
&\norm{\Rlqr\inv B\tp P(\epsilon)^{\frac{1}{2}}v}^2\leq \frac{\bound^2}{\ell_P}\norm{v}^2 &  &\forall v\in\R 6.
\end{aligned}
\end{equation}
From Proposition~9.4.9 in~\cite{bernstein_matrix_2009}, it follows that~\eqref{eq:svmaxPeps} is satisfied if 
\begin{equation}
\label{eq:svmaxP}
\begin{aligned}
\svmax(\Rlqr\inv B\tp P(\epsilon)^{\frac{1}{2}})^2\leq\frac{\bound^2}{\ell_P}
\end{aligned}
\end{equation}
holds. Since $P(\epsilon)$ can be made arbitrarily close to zero by choosing a small enough $\epsilon$, it follows that,  for each $\ell_P>0$, $\ell_H>0$, $\Rlqr\vg 0$ and $\bound>0$, it is possible to find $P(\epsilon)$ satisfying both~\eqref{eq:svmaxP} and~\eqref{eq:PH}.
\qed\end{pf}

The previous proposition is {very important} to the following theorem, which constitutes the main result of this section.

\begin{theorem}\label{thm:pos}
For each $\bound\in (0,\norm{g}-M_2)$, and each compact set $U\subset\R 6$, there exists $K\in\R{3\x 6}$ such that the origin of the closed-loop system resulting from the interconnection between~\eqref{eq:doubleint} and~\eqref{eq:w} is exponentially stable from $U$. Moreover, each solution to~\eqref{eq:doubleint} from $U$, denoted by $t\mapsto (\pe,\ve)(t)$,  satisfies 
$\norm{K\bmtx{\pe(t)\tp & \ve(t)\tp}\tp}\leq \bound$
for each $t\geq 0$.
\end{theorem}
\begin{pf}
 Choosing a positive definite matrix $H\in\defpos{6}$ such that $\sublevel{H}(1)$ is a bounding ellipsoid for $U$, it follows from Proposition~\ref{pro:ineqs} that there exists $K\in\R{3\x 6}$ and $\Qlqr,P\in\defpos{6}$ such that the conditions~\eqref{eq:Hellip} and~\eqref{eq:QR} hold for any positive definite matrix $\Rlqr\in\R{3\x 3}$ and any $\ell_P>0$.
Let $V_p(\pe,\ve)\ceq\bmtx{\pe\tp & \ve\tp} P\bmtx{\pe\tp& \ve\tp}\tp,$
for each $(\pe,\ve)\in\R 6$, which is a positive definite function relative to $\{(\pe,\ve)\in\R 6:\pe=\ve=0\}$ and satisfies: 
\IfFull{%
\begin{equation}
\label{eq:dVp}
\begin{multlined}
\inner{\grad V_p(\pe,\ve)}{f_p(\pe,\ve)}\\ %&=\bmtx{\pe \\ \ve}\tp \left((A+BK)\tp P+P(A+BK)\right)\bmtx{\pe\\ \ve}\\
\leq -\bmtx{\pe\tp & \ve\tp} (\Qlqr+K\tp \Rlqr K)\bmtx{\pe\tp & \ve\tp}\tp
\end{multlined}
\end{equation}
}{$\inner{\grad V_p(\pe,\ve)}{f_p(\pe,\ve)}\leq -\bmtx{\pe\tp & \ve\tp} (\Qlqr+K\tp \Rlqr K)\bmtx{\pe\tp & \ve\tp}\tp,$}
where $f_p(\pe,\ve)\ceq(\ve, K\bmtx{\pe\tp & \ve\tp}\tp)$ for each $(\pe,\ve)\in\sublevel{P}(\ell_P).$
It follows from the conditions~\eqref{eq:Hellip}, that every solution $t\mapsto (\pe,\ve)(t)$ to~\eqref{eq:doubleint} from $U$ satisfies $\norm{K\bmtx{\pe(t) \\ \ve(t)}}\leq\bound< \norm{g}-M_2$ for all $t\geq 0$.
\qed\end{pf}

\IfFull{
In the next lemma, we show that the controller gain $K$ in~\eqref{eq:w} can be computed from an optimization problem that provides a sub-optimal solution to the $H_2$-minimization problem (c.f. Proposition~3.11 in~\cite{scherer_linear_2000}).
\begin{lemma}\label{lem:LMIs}
Given $\ell_H>0$, $\ell_P>0$, $\Qlqr\in\defpos{6}$ and $\Rlqr\in\defpos{3}$, there exists a solution $P\in\defpos{6}$, $K\in\R{3\x 6}$ to~\eqref{eq:Hellip} and~\eqref{eq:QR} if and only if there exists a solution $Y\in\defpos{6}$ and $L\in\R{3\x 6}$ to 
\begin{equation}
\label{eq:H2}
\begin{array}{cc}
\text{minimize} & \trace(Y\inv)\\
\text{subject to} & (Y, L)\in \chi_{LMI}
\end{array}
\end{equation}
where $\chi_{LMI}$ is the set of linear matrix inequalities:
\begin{subequations}
\mathtoolsset{showonlyrefs=false}
\label{eq:chilmi}
\begin{align}
\label{eq:chilmiQR}&\bmtx{-(AY+BL)\tp-(AY+BL) & Y & L\tp\\
Y & \Qlqr\inv & 0\\
L & 0 & \Rlqr\inv}\vgeq 0\\
\label{eq:YH}&Y\vgeq\frac{\ell_H}{\ell_P} H\inv\\
\label{eq:YL}&\bmtx{Y & L\\ L\tp & \frac{\bound^2}{\ell_P}\eye{6}}\vgeq 0
\end{align}
\end{subequations}
in which case $P=Y\inv$ and $K=LY\inv$.
\end{lemma}
\begin{pf}
Since $\Qlqr\vg 0$, we can rewrite~\eqref{eq:QR} as follows:
\begin{equation}
\label{eq:cond2}
\begin{aligned}
\bmtx{X-Y\Qlqr Y-L\tp \Rlqr L &0\\
0 & \Qlqr\inv}\vgeq 0
\end{aligned}
\end{equation}
where $X\ceq -(AY+BL)\tp-(AY+BL)$.
Equivalently, we obtain
\begin{equation}
\label{eq:cond3}
\begin{aligned}
\bmtx{X & Y\\ Y & \Qlqr\inv}-\bmtx{L\tp\\ 0}\Rlqr\bmtx{L & 0}\vgeq 0
\end{aligned}
\end{equation}
from~\eqref{eq:cond2} by means of a congruence transformation. The application of Proposition~8.2.4.iii in~\cite{bernstein_matrix_2009} to~\eqref{eq:cond3} yields~\eqref{eq:chilmiQR}. The condition~\eqref{eq:YH} follows from the fact that $\sublevel{H}(\ell_H)\subset \sublevel{P}(\ell_P)$ if and only if $P\vleq \frac{\ell_P}{\ell_H} H,$ using $Y=P\inv$. The condition~\eqref{eq:YL} follows from the fact that $\sublevel{P}(\ell_P)\subset\sublevel{K\tp K}(\bound^2)$ if and only if $\frac{K\tp K}{\bound^2}\vleq \frac{P}{\ell_P}$ and Proposition~8.2.4 in~\cite{bernstein_matrix_2009} as follows:
 \begin{equation}
 \begin{aligned}
  \frac{K\tp K}{\bound^2}\vleq \frac{P}{\ell_P}&\iff \frac{YK\tp K Y}{\bound^2}\vleq \frac{Y}{\ell_p}\\
  &\iff \frac{Y}{\ell_P}-\frac{L\tp L}{\bound^2}\vgeq 0
 \end{aligned}
 \end{equation}
where we have, once again, used $Y=P\inv$ and $L=K Y$.
\end{pf}
}{}%endif

\subsection{Partial attitude tracking}\label{sec:partial}
In this section, we develop a controller for~\eqref{eq:quad} that tracks a reference trajectory satisfying Assumption~\ref{ass:pd}. The desired acceleration, imposed by the reference trajectory upon the vehicle, is achieved by aligning the thrust vector $\Rmat r$ with the direction of the desired acceleration. We refer to this as partial attitude tracking because we do not control rotations around the thrust vector. 
\IfHub{%
Given $k>0$ and $\yrbnd\in(-1,1)$, we define $$V(x,y)\ceq\frac{1-r\tp x}{1-r\tp x+k(1-x\tp y)}$$
for each $(x,y)\in\sphere 2\x\yset$, where $\yset\ceq\{y\in\sphere 2: y\tp r\leq\yrbnd\}.$ Defining 
\begin{subequations}
\mathtoolsset{showonlyrefs=false}
\label{eq:aagg}
\begin{align}
\label{eq:laa}\lboundVone&=\frac{1}{2(1+k+\sqrt{1+2k\yrbnd+k^2})},\\
\label{eq:uaa}\uboundVone&=\frac{1}{2(1+k-\sqrt{1+2k\yrbnd+k^2})},\\
\label{eq:gg}\ubounddVone&=\frac{2k(1-\maxVflow)(1-\yrbnd)}{\left(1+k+\sqrt{1+2k\yrbnd+k^2}\right)^2}.
\end{align}
\end{subequations}
and 
\begin{subequations}
\label{eq:C1D1}
\mathtoolsset{showonlyrefs=false}
\begin{align}
\label{eq:C1} C&\ceq\{(x,y)\in\sphere 2\x\yset: \gap(x,y)\leq\delta\},\\
\label{eq:D1} D&\ceq\{(x,y)\in\sphere 2\x\yset: \gap(x,y)\geq\delta\},
\end{align}
\end{subequations}
where $\gap(x,y)\ceq V(x,y)-\min\{V(x,,y):y\in\yset\}$ for each $(x,y)\in\sphere 2\x\yset$ and $\maxVflow\ceq\max\{V(x,y):(x,y)\in C\}\leq \delta+\frac{2}{2+k(1+\gamma)}$, it is possible to verify that
\begin{equation}
\label{eq:V1bounds}
\begin{aligned}
&\lboundVone\norm{x-r}^2\leq V(x,y)\leq\uboundVone\norm{x-r}^2 & &
\forall (x,y)\in C\cup D= \sphere 2\x\yset\\
&\norm{\PTSn(x)\grad V^{y}(x)}^2\geq\ubounddVone V(x,y) & &
\forall (x,y)\in C.
\end{aligned}
\end{equation}
Using the previous construction, we define the
}{%
For controller design purposes, let us assume the following.

\begin{assumption}\label{ass:Vone}
Given $\delta>0$, there exists a function $V\in\spf[r]{\yset}$ with synergy gap exceeding $\delta$ that satisfies
\begin{equation}
\label{eq:V1bounds}
\begin{aligned}
\lboundVone\norm{x-r}^2\leq V(x,y)&\leq\uboundVone\norm{x-r}^2 &
&\forall (x,y)\in C\cup D\\
\norm{\PTSn(x)\grad V^{y}(x)}^2&\geq\ubounddVone V(x,y) &
& \forall (x,y)\in C.
\end{aligned}
\end{equation}
for some $\lboundVone,\uboundVone,\ubounddVone>0$, where 
\begin{subequations}
\label{eq:C1D1}
\mathtoolsset{showonlyrefs=false}
\begin{align}
\label{eq:C1} C&\ceq\{(x,y)\in\sphere 2\x\yset: \gap(x,y)\leq\delta\},\\
\label{eq:D1} D&\ceq\{(x,y)\in\sphere 2\x\yset: \gap(x,y)\geq\delta\}.
\end{align}
\end{subequations}
\end{assumption}

Under Assumption~\ref{ass:Vone}, the application of the controller developed in Section~\ref{sec:synergistic} yields
}%endif
 the closed-loop system $\Hmc_1\ceq (C_1,F_1,D_1,G_1)$ with state $\zeta\ceq(z,\Rmat,y)\in \Zmc\ceq \trajset\x\R 6\x\SO(3)\x\yset$, given by
\IfHub{
\begin{equation}
\label{eq:qcloop1}
\begin{aligned}
&\begin{multlined}[0.79\textwidth]
F_1(\zeta)\ceq\left\{\pmtx{F_p(p_d^{(3)},\zeta)\\ \Rmat\sk(\wctrl{1}(p_d^{(3)},\zeta)) \\0}:p_d^{(3)}\in M_3\ball\right\} \\
\forall\zeta\in C_1\ceq\{\zeta\in\Zmc:(\Rmat\tp\Td(z),{y})\in C\}
\end{multlined}\\
&G_1(\zeta)\ceq \pmtx{ z\\ \Rmat\\ \yminV(\Rmat\tp\Td(z))}\ 
\forall\zeta\in D_1\ceq\{\zeta\in\Zmc:(\Rmat\tp \Td(z),{y})\in D\}
\end{aligned}
\end{equation}
}{
\begin{equation}
\label{eq:qcloop1}
\begin{aligned}
F_1(\zeta)\ceq\{(F_p(p_d^{(3)},\zeta),\Rmat\sk(\wctrl{1}(p_d^{(3)},\zeta)),0):p_d^{(3)}\in M_3\ball\}&\\
\zeta\in C_1\ceq\{\zeta\in\Zmc:(\Rmat\tp\Td(z),{y})\in C\}&\\
\begin{multlined}[0.45\textwidth]
G_1(\zeta)\ceq(z,\Rmat,\yminV(\Rmat\tp\Td(z)))\\
\zeta\in D_1\ceq\{\zeta\in\Zmc:(\Rmat\tp \Td(z),{y})\in D\}
\end{multlined}&
\end{aligned}
\end{equation}
}%endif
where \IfHub{$\yminV(x)\ceq\argmin\{V(x,y):y\in\yset\}$ for each $x\in\sphere 2$,}{}
%
%\begin{equation}
%\label{eq:Fp}
%\begin{aligned}
$F_p(p_d^{(3)},\zeta)\ceq (p_d^{(3)}, \ve, \Rmat r \uctrl(z,\Rmat)+\grv-\ddot p_d)$
%\end{aligned}
%\end{equation}
for each $(p_d^{(3)},\zeta)\in M_3\ball\x\Zmc$, the inputs $u$ and $\wb$ were assigned to $\uctrl(z,\Rmat)$ and 
\begin{equation}
\label{eq:wctrl1}
\begin{multlined}
\wctrl{1}(p_d^{(3)},\zeta)\ceq\sk(\Rmat\tp\Td(z))\big(\Rmat\tp \D[z]{\Td(z)} F_p(p_d^{(3)},\zeta)\\
+(k_1+k_p\adapt(z))\grad V^{{y}}(\Rmat\tp\Td(z))\big),
\end{multlined}
\end{equation}
for each $(p_d^{(3)},\zeta)\in M_3\ball\x\Zmc$, respectively, with $k_1>0$, $k_p>0$ and 
\begin{equation}
\label{eq:adapt}
\begin{aligned}
\adapt(z)\ceq \frac{2}{\sqrt{\lboundVone}}\svmax\left(\bmtx{0& \eye{3}}P^{\frac{1}{2}}\right)\norm{w(\pe,\ve)-\grv+\ddot p_d},
\end{aligned}
\end{equation}
for each $z\in \trajset\x\R 6$. \IfHub{}{The closed-loop system represented by $\Hmc_1$ in~\eqref{eq:qcloop1} inherits the switching logic from~\eqref{eq:controller} but not the same feedback law. The feedback law~\eqref{eq:wctrl1} is comprised of negative feedback of the attitude tracking error as in~\eqref{eq:controller}, but also a feedforward term that takes into account the fact that the reference we wish to track is not a constant. Moreover, the function $\adapt$ is used to increase the gain of the attitude controller as the position error increases.}
Given a reference trajectory satisfying Assumption~\ref{ass:pd}\IfHub{ and a}{,} compact set of initial conditions $U\subset\R 6$ for the position and velocity errors \IfHub{}{and a synergistic potential function satisfying Assumption~\ref{ass:Vone}, }the controller design is as follows:
\begin{enumerate}[label=(C\arabic*)]
\item \label{ass:C1}Select $H\in\defpos{6}$ such that $\sublevel{H}(1)$ is a bounding ellipsoid for $U$;
\item \label{ass:C2} Select $k_p,\bar k_1>0$ so that 
\begin{equation}
\label{eq:bark1}
\begin{aligned}
k_p \bar k_1\ubounddVone > 1
\end{aligned}
\end{equation}
\IfHub{}{where $\ubounddVone$ is given in Assumption~\ref{ass:Vone};}
\item \label{ass:C3} {Given $\bound\in (0,\norm{\grv}-M_2)$ and $\Rlqr\in\defpos{6}$, select $\ell_H=\ell_P=(1+\bar\nu)^2,$ where 
\begin{equation}
\label{eq:barnu}
\begin{aligned}
\bar\nu\ceq\bar k_1\sqrt{\max_{(x,{y})\in\sphere 2\x\yset}V(x,{y})}
\end{aligned}
\end{equation}
and $\Qlqr\in\defpos{6}$ small enough and so that the constraints~\eqref{eq:Hellip} and~\eqref{eq:QR} are feasible;}
\item \label{ass:C4}Compute the controller gain $K$ by means of the optimization problem \IfHub{in Lemma~\ref{lem:LMIs}.}{$\min\{\trace(P):(P,K)\in\chi\}$ where $\chi$ represents the constraints~\eqref{eq:Hellip} and~\eqref{eq:QR}.}
\end{enumerate}
This controller design enables exponential stability, as proved next.

\begin{theorem}\label{thm:quad1}
Let Assumption\IfHub{}{s}~\ref{ass:pd}\IfHub{}{ and~\ref{ass:Vone}} hold. For each compact set $U\subset\R 6$, if~\ref{ass:C1}--\ref{ass:C4} are satisfied, then $\Amc_1\ceq \{\zeta\in\Zmc:\pe=\ve =0,\Td(z)=\Rmat r\}$ is exponentially stable in the $t$-direction from $\trajset\x U\x\SO(3)\x\yset$ for the hybrid system~\eqref{eq:qcloop1}.
\end{theorem}
\begin{pf}
\IfHub{}{Firstly, we prove that the hybrid system~\eqref{eq:qcloop1} satisfies the hybrid basic conditions. It follows from Proposition~\ref{prop:muIsNice} that $\gap$ is continuous, thus both $C_1$ and $D_1$ are closed because they correspond to the inverse image of a closed set through a continuous map. It follows from $\bound\in(0,\norm{g}-M_2)$ and the properties of the saturation function that $\D[z]{\Td(z)}$ is continuously differentiable for each $z\in \trajset\x\R 6$, hence $F_1$ is outer semicontinuous and locally bounded relative to $C_1$. It follows from the outer semicontinuity of $\yminV$ that $G_1$ is outer semicontinuous relative to $D_1$ and, since $\yminV$ takes values over a compact set, $G_1$ is locally bounded relative to $D_1$.}
Let 
\begin{equation}
\label{eq:W1}
\begin{aligned}
W_1(\zeta)&\ceq \sqrt{V_p(\pe,\ve)}+\bar k_1\sqrt{V(x,{y})} & &\forall\zeta\in\Zmc
\end{aligned}
\end{equation}
with $P:=Y\inv$ and $x\ceq\Rmat\tp\Td(z)$. \IfHub{We have}{It follows from Assumption~\ref{ass:Vone}} that 
\begin{equation}
\label{eq:W1bounds}
\begin{multlined}
\min\{\sqrt{\eigmin(P)},\bar k_1\sqrt{\lboundVone}\}\norm{(\pe,\ve,x-r)} \leq W_1(\zeta)\\
\leq \sqrt{\eigmax(P)+\bar k_1^2\uboundVone}\norm{(\pe,\ve,x-r)}.
\end{multlined}
\end{equation}
for each~$\zeta\in C_1\cup D_1$. It follows from the assumptions that
\IfFull{%
the time derivative of~\eqref{eq:W1} is given by 
\begin{equation}
\label{eq:dW1_0}
\begin{aligned}
&\inner{W_1(\zeta)}{f_1}=\\
&+\frac{1}{2\sqrt{V_p(\pe,\ve)}}\grad V_p(\pe,\ve)\tp \bmtx{\ve\\ \Rmat r\uctrl(z,\Rmat)+\grv-\ddot p_d}\\
&+\frac{\bar k_1}{2\sqrt{V(x,y)}}\grad V^{y}(x)\tp (\sk(x)\wctrl{1}(\zeta,y)\\
&-\Rmat\tp\D[z]{\rho(z)}F_p(p^{(3)}_d,\zeta))
\end{aligned}
\end{equation}
for each $f_1\in F_1(\zeta)$, $\zeta \in \hat\Omega\ceq\{\zeta\in\Zmc: W_1(\zeta)\leq 1+\bar\nu\}$, because $W_1(\zeta)\leq 1+\bar\nu$ implies that $V_p(\pe,\ve)\leq (1+\bar\nu)^2$ and, by the construction~\ref{ass:C1}-\ref{ass:C3}, the saturation bound $\bound$ is not exceeded in this set since it satisfies~\eqref{eq:Hellip}. Using $x x\tp =\sk(x)^2+\eye{3}$ for each $x\in\sphere 2$, and replacing~\eqref{eq:ku} and~\eqref{eq:wctrl1} into~\eqref{eq:dW1_0} yields
\begin{equation}
\label{eq:dW1_1}
\begin{aligned}
& W_1\ggrad(\zeta;f_1)=\frac{1}{2\sqrt{V_p(\pe,\ve)}}\grad V_p(\pe,\ve)\tp \bmtx{\ve\\ w(\pe,\ve)}\\
&+\frac{1}{2\sqrt{V_p(\pe,\ve)}}\grad V_p(\pe,\ve)\tp \bmtx{0\\ \eye{3}}\sk(\Rmat r)^2(w(\pe,\ve)-\grv+\ddot p_d)\\
&-\frac{\bar k_1(k_1+k_p\adapt(z))}{2\sqrt{V(x,y)}}\norm{ \sk(x)^2\grad V^{y}(x)}^2
\end{aligned}
\end{equation}
for each $f_1\in F_1(\zeta)$, $\zeta\in\hat\Omega$. Using $\grad V_p(\pe,\ve)=2 P\bmtx{\pe\tp & \ve\tp}\tp$ and~\eqref{eq:dVp} it follows from~\eqref{eq:dW1_1} that
\begin{equation}
\label{eq:dW1_2}
\begin{aligned}
& W_1\ggrad(\zeta;f_1)\leq-\frac{1}{2\sqrt{V_p(\pe,\ve)}}\bmtx{\pe\\ \ve}\tp(\Qlqr+K\tp\Rlqr K) \bmtx{\pe\\ \ve}\\
&+\frac{1}{\sqrt{V_p(\pe,\ve)}}\norm{ \bmtx{0 & \eye{3}}P \bmtx{\pe\\ \ve}}\norm{\sk(\Rmat r)^2(w(\pe,\ve)-\grv+\ddot p_d)}\\
&-\frac{\bar k_1(k_1+k_p\adapt(z))}{2\sqrt{V(x,y)}}\norm{ \sk(x)^2\grad V^{y}(x)}^2
\end{aligned}
\end{equation}
for each $f_1\in\ F_1(\zeta)$ $\zeta\in \hat\Omega$. Since 
\begin{equation}
\norm{\sk(\Rmat r)^2(w(\pe,\ve)-\grv+\ddot p_d)}
\leq \norm{x-r}\norm{w(\pe,\ve)-\grv+\ddot p_d}
\end{equation}
for each $\zeta \in C_1\cup D_1$ and 
\begin{equation}
\label{eq:Pineq}
\begin{aligned}
\frac{\norm{\bmtx{0 &\eye{3}} P z}}{\sqrt{z\tp P z}}&=\frac{\norm{\bmtx{0 &\eye{3}} P^{1/2} \tilde z}}{\norm{\tilde z}}
&\leq \svmax\left(\bmtx{0 &\eye{3}} P^{1/2}\right)
\end{aligned}
\end{equation}
for each $z\in\R 6\minus\{0\}$ where $\tilde z\ceq P^{1/2}z$, it follows~\eqref{eq:dW1_2} that 
\begin{equation}
\label{eq:dW1_3}
\begin{aligned}
& W_1\ggrad(\zeta;f_1)\leq -\frac{\eigmin(\Qlqr+K\tp\Rlqr K)}{2\sqrt{\eigmax(P)}}\sqrt{V_p(\pe,\ve)}\\
&+\svmax \left(\bmtx{0 & \eye{3}}P^{1/2}\right)\norm{x-r}\norm{w(\pe,\ve)-\grv+\ddot p_d}\\
&-\frac{\bar k_1(k_1+k_p\adapt(z))}{2\sqrt{V(x,y)}}\norm{ \sk(x)^2\grad V^{y}(x)}^2
\end{aligned}
\end{equation}
for each $f_1\in F_1(\zeta)$, $\zeta\in \hat\Omega$.  Replacing~\eqref{eq:adapt} into~\eqref{eq:dW1_3}, we have that
\begin{equation}
\label{eq:dW1_4}
\begin{aligned}
 W_1\ggrad(\zeta;f_1)&\leq-\frac{\eigmin(\Qlqr+K\tp\Rlqr K)}{2\sqrt{\eigmax(P)}}\sqrt{V_p(\pe,\ve)}\\
&-\frac{\bar k_1(k_1+k_p\adapt(z))\ubounddVone}{2}\sqrt{V(x,y)}\\
&+\frac{\adapt(z)}{2}\sqrt{V(x,y)}
\end{aligned}
\end{equation}
for each $f_1\in F_1(\zeta)$, $\zeta\in \hat\Omega$. It follows from~\ref{ass:C2} that
}{}%endif
\begin{equation}
\label{eq:dV1}
\begin{aligned}
 W_1\ggrad(\zeta;f_1)&\leq -\frac{\eigmin(\Qlqr+K\tp \Rlqr K)}{2\sqrt{\eigmax(P)}}\sqrt{V_p(\pe,\ve)}\\
&-\frac{\bar k_1 k_1\ubounddVone \sqrt{V(x,{y})}}{2}
\end{aligned}
\end{equation}
for each $f_1\in F_1(\zeta)$, \IfHub{$\zeta\in \hat\Omega$}{$\zeta\in \hat\Omega\ceq\{\zeta\in\Zmc: W_1(\zeta)\leq 1+\bar\nu\}$}. {It follows from~\eqref{eq:Hellip} that $(\pe,\ve)\in\sublevel{H}(1)$ implies $(\pe,\ve)\in\sublevel{P}(1)$, hence,} for each solution from $\trajset\x U\x\SO(3)\x\yset$, we have $V_p(\pe(0,0),\ve(0,0))\leq 1$ and, consequently, $W_1(\zeta(0,0))\leq 1+\bar\nu$. From~\eqref{eq:bark1} and~\eqref{eq:dV1}, we have $W_1\ggrad(\zeta;f_1)\leq -\lambda W_1(\zeta)$ for each $f_1\in F_1(\zeta)$, $\zeta\in \hat\Omega$ with $\lambda\ceq\min\left\{\frac{\eigmin(\Qlqr+K\tp \Rlqr K)}{2\sqrt{\eigmax(P)}},\frac{\bar k_1 k_1\ubounddVone }{2}\right\}.$
Note that $V$ is a synergistic potential function relative to $r$ by assumption, thus it satisfies $V(x,g_{y})\leq V(x,y)-\delta$ for each $g_{y}\in\yminV(x)$ and $\zeta\in D_1$ by construction, and $G_1(D_1)\cap D_1=\emptyset$ because $\gap(x,g_{y})=0<\delta$ for each $g_{y}\in \yminV(x)$ for each $\zeta\in D_1$. {Each solution $(t,j)\mapsto\zeta(t,j)$ to~\eqref{eq:qcloop1} from $\trajset \x U\x\SO(3)\x\yset$ is such that the initial condition $\zeta(0,0)$ belongs to the compact set $\hat\Omega$. Since $W_1$ is strictly decreasing during flows and jumps of~\eqref{eq:qcloop1}, it follows that $\hat\Omega$ is forward invariant for each solution from $\trajset \x U\x\SO(3)\x\yset$.
Since $G_1(D_1)\subset C_1\cup D_1$ and $F_1(\zeta)$ belongs to the tangent cone to $C_1$ at $\zeta$ for each $\zeta\in C_1\minus D_1$, it follows from Proposition~6.10 in~\cite{goebel_hybrid_2012} that each maximal solution to $\Hmc_1$ from $\trajset\x U\x\SO(3)\x\yset$ is complete, hence $\Amc_1$ is exponentially stable in the $t$-direction from $\trajset \x U\x\SO(3)\x\yset$ for the hybrid system~\eqref{eq:qcloop1}.}
\qed\end{pf}

It should be pointed out that, underlying the controller design described in~\ref{ass:C1} through~\ref{ass:C4}, there is a trade-off to be resolved: if $k_p<< 1$, then the position controller is going to have a low gain $K$ which results in large deviations from the reference; on the other hand, if controller gain $K$ is large, the function $\adapt$ might increase the gain on the attitude controller beyond what is acceptable in practical terms.

\subsection{Simulation Results}\label{sec:sim}
\begin{figure*}
\centering
\psfragscanon
\input{flowfields.tex}
\includegraphics[width=0.8\textwidth]{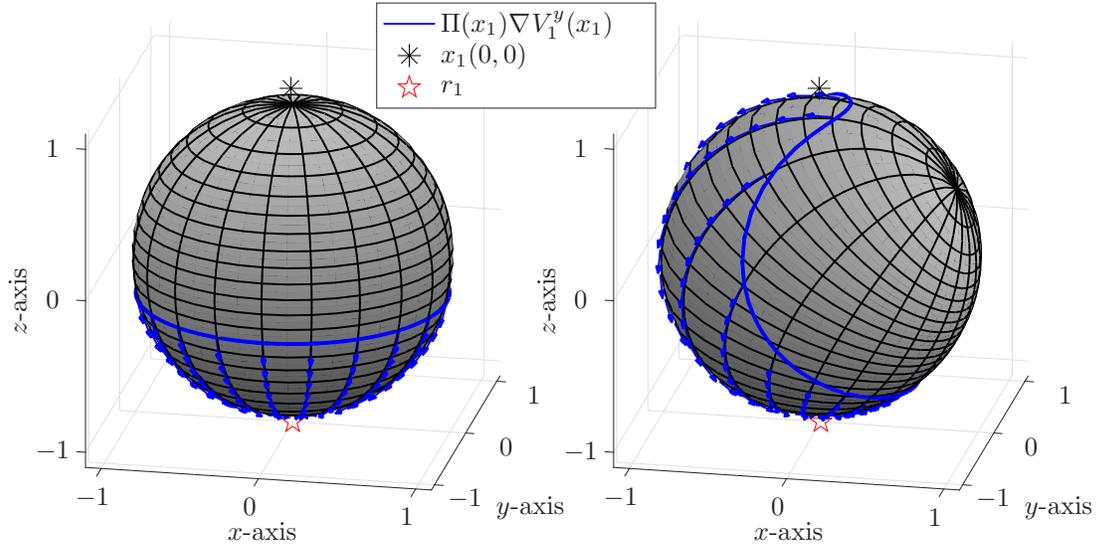}
\caption{The arrows represent the gradient vector field of $x_1\to V_1^{y_1}(x_1)$ with $r_1=[0\ 0\ -1]	\tp$ and $x_1\ceq\Rmat\rho_1(z)$ restricted to the set $\{x_1\in\sphere 2: \gap(x_1,{y_1})\leq\delta\}$ for two different values of ${y_1}$: on the left we have ${y_1}=[0\ 0\ 1]\tp$ and on the right we have ${y_1}=[\sqrt{3}/2\   0\    0.5]\tp $. The value $x_1(0,0)\approx[-0.0101\         0\    0.9999]\tp$ lies outside the flow set when the former value of ${y_1}$ is used, generating a jump of the logic variable to the latter value.}
\label{fig:flowfields}
\hrulefill
\end{figure*}

\IfFull{%
\begin{equation}
\label{eq:initx}
\begin{aligned}
p(0,0)&\ceq p_d(0,0), & v(0,0)&\ceq v_d(0,0), \\
 {y_1}(0,0)&=\bmtx{0&   0 &  1}\tp, & \Rmat(0,0)&\ceq\diag\left(\bmtx{1& -1 & -1}\right).
\end{aligned}
\end{equation}
The initial conditions are such that the initial position and velocity tracking errors are zero and the vehicle is upside down, that is, the thrust vector is pointing in the direction of the gravity vector. As a result, the vehicle needs to quickly reorient itself in order to track the desired reference trajectory. To achieve this, the initial orientation error $x_1(0,0)$ triggers the update of the memory variable $y_1$ that is represented in Figure~\ref{fig:flowfields}. It is possible to verify in Figure~\ref{fig:sim_Re} that the controller switching (which occurs at $t=0$) leads to an instantaneous decrease of the potential function $V_1$ and, additionally, the function is strictly decreasing as a function of continuous time, corresponding to a fast attitude correction within the first few seconds of the simulation. 
While the vehicle reorients itself to track the desired thrust vector, the position and the velocity tracking errors increase but this transient response is soon mitigated and the vehicle converges to the desired trajectory, as can be verified in Figure~\ref{fig:sim}.
%It is possible to verify that the orientation converges much faster than the position and velocity errors, which is to be expected. It is also possible to verify that despite the very distinct yaw motion, the two controllers (with and without yaw stabilization) behave similarly in terms of position and velocity, which illustrates the decoupling between the yaw error and the remaining states. In Figure~\ref{fig:flows}, it is possible to compare the evolution of $x_1$ and $x_2$ with and without yaw stabilization. The evolution of $x_1$ is the same in both cases, while the evolution of $x_2$ is widely different. When yaw stabilization is active, the state $x_2$ converges to $r_2$ and when it is not, $x_2$ converges to to a vector that is orthogonal to $r_1$. Coincidentally, for the given initial conditions, $x_2$ converges to a vector close to $-r_2$. 
\begin{figure}
\centering
\input{sublevels.tex}
\includegraphics[width=0.45\textwidth]{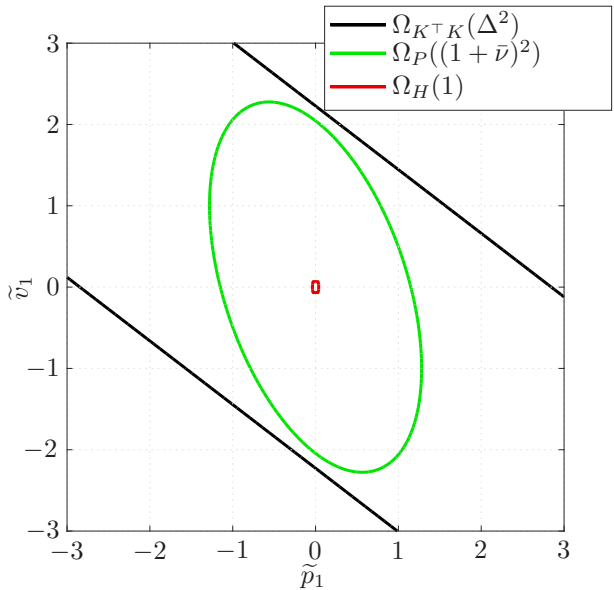}
\caption{Representation of a cross-section of the sets $\sublevel{H}(1)$, $\sublevel{P}(\ell_P)$, $\sublevel{K\tp K}(\Delta^2)$ and of the solution to the closed-loop hybrid system for the controller parameters and initial conditions presented in Section~\ref{sec:experiments}. Solutions with initial conditions in $\sublevel{H}(1)$ (red inner circle) do not leave $\sublevel{P}((1+\bar\nu)^2)$ (green outer ellipse), hence do not violate the saturation constraints, which would happen if solutions were to cross the dark parallel lines.}
\label{fig:sublevels}
\end{figure}
}{}

\IfFull{%
\begin{figure*}
\centering
\input{sim.tex}
\includegraphics[width=0.9\textwidth]{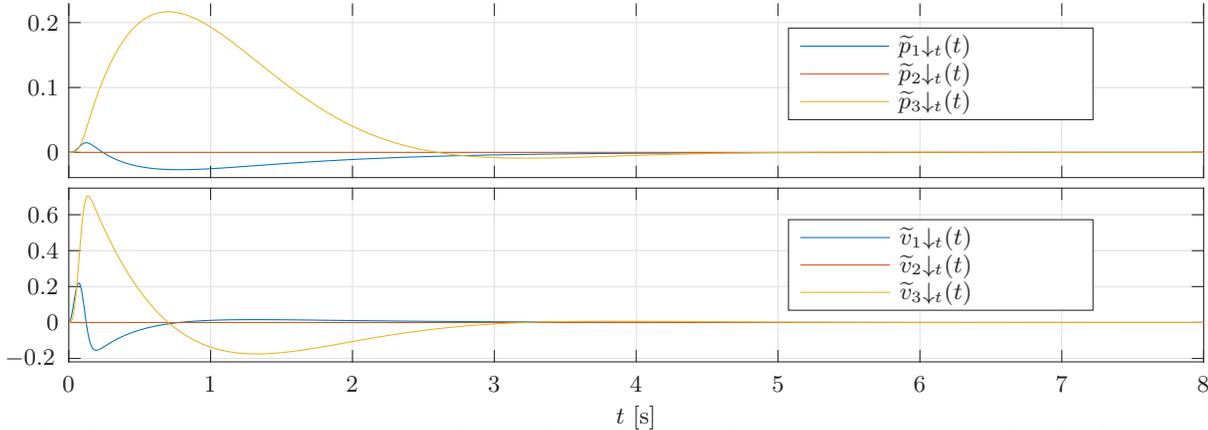}
\caption{Simulation results representing the evolution of the position and velocity tracking errors for the closed-loop system with initial conditions and controller parameters given in Section~\ref{sec:experiments}.}
\label{fig:sim}
\hrulefill
\end{figure*}

\begin{figure}
\centering
\input{sim_Re.tex}
\includegraphics[width=0.5\textwidth]{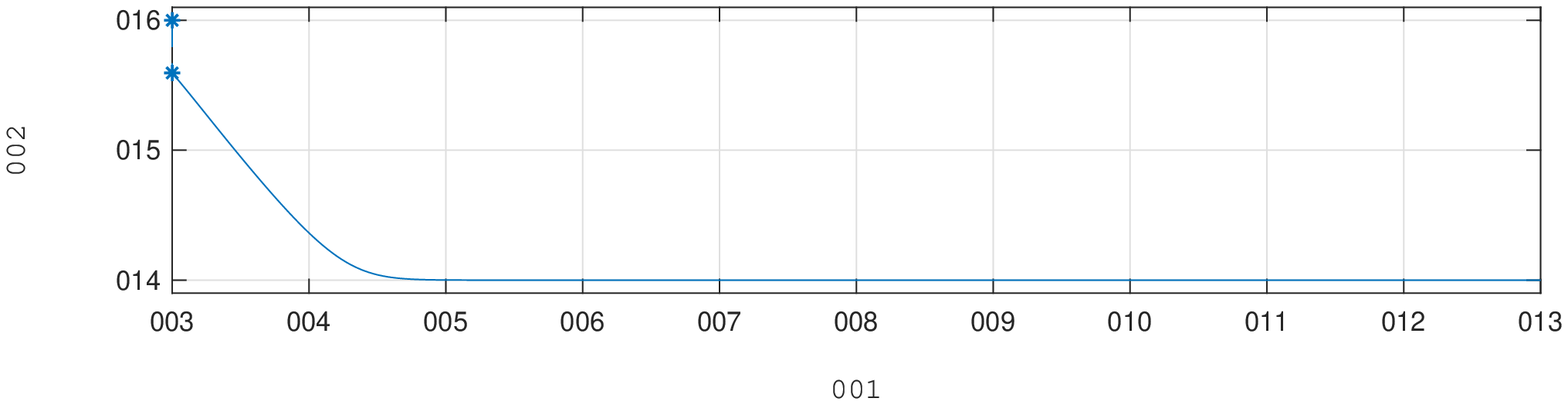}
\caption{Simulation results representing the evolution of the potential function $t\mapsto V_1\projt(t)\ceq V_1(\Rmat\projt(t)\tp\Td(z\projt(t)),y_1\projt(t))$ for the closed-loop system with initial conditions and controller parameters given in Section~\ref{sec:sim}.}
\label{fig:sim_Re}
\end{figure}
}{}%endif

%\begin{figure}
%\centering
%\psfragscanon
%%\input{flows.tex}
%%\includegraphics[width=0.9\textwidth]{flows.eps}
%\caption{Representation of the trajectories of $x_1$ and $x_2$ with and without yaw stabilization, for the parameters of Section~\protect\ref{sec:sim}.}
%\label{fig:flows}
%\end{figure}

\IfFull{%
\subsection{Experimental Results}\label{sec:experiments}
In this section, we present the experimental results for quadrotor trajectory tracking using the controller designed in Section~\ref{sec:partial} with the synergistic potential function~\eqref{eqn:V}. 
Given $f>0$, the reference trajectory is given by $t\mapsto p_d(t)=\bmtx{\cos 2\pi ft & \sin 2\pi ft & 0}\tp$ for each $t\geq 0$. The reference acceleration and jerk, given by $\ddot p_d$ and $p^{(3)}_d$, respectively, satisfy Assumption~\ref{ass:pd} with $M_2=(2\pi f)^2$ and $M_3=(2\pi f)^3$. The thrust vector is $r_1=\bmtx{0 & 0 &-1}\tp$, the gravity vector is given by $\grv=\bmtx{0 & 0 & 9.81}\tp$. 
Assumption~\ref{ass:Vone} is satisfied with $\lboundVone,$ $\uboundVone$ and $\ubounddVone$ given by~\eqref{eq:laa},~\eqref{eq:uaa} and~\eqref{eq:gg}, respectively. For $k=1,$ $\yrbnd=-0.5,$ $\delta=\frac{1+\yrbnd}{2\left(\frac{2}{k}+1+\yrbnd\right)}=0.1$ we obtain $\uboundVone=0.5$, $\lboundVone=1/6,$ $\maxVflow\leq \frac{2}{2+k(1+\yrbnd)}+\delta=0.82,$ and $\ubounddVone=0.18$.
We design the position controller following~\ref{ass:C1} through~\ref{ass:C4} considering $U\ceq\sublevel{H}(1)$ with $\bar k_1 =12$, $k_p=1$ and 
\begin{equation}
\label{eq:LQR}
\begin{aligned}
H&\ceq\diag\left(\bmtx{500 & 500 & 500 & 100 & 100 & 100}\right),\\ 
\Qlqr&\ceq\diag\left(\bmtx{10 & 10 & 100 & 100 & 100 & 1}\right),\\ 
\Rlqr&\ceq10\eye{3}.
\end{aligned}
\end{equation}
%
%$\ttset$ given in Example~\ref{exa:vfield} and we make use of the stereographic projection to define a chart on $\sphere 2\minus\{-r_1\}$.%, which is represented in Figure~\ref{fig:stereo} for the initial conditions $(y,\tt)=(y_0,\alpha_0,\alpha_\ell,\ell)=(\bmtx{0.3956 &  -0.6332 &    0.6652}\tp,\pi/2,0,1)$, where $y_0\in\sphere 2$ was randomly selected from the set $\yset\ceq\{y\in\sphere 2: y\tp r_1\leq\yrbnd\}$. 

%\begin{figure}
%\centering
%\input{stereo.tex}
%\includegraphics[width=0.5\textwidth]{stereo.eps}
%\caption{Representation of $\vfield{(y,\tt)}$ with $y=[0.3956\   -0.6332\     0.6652]\tp$ and $\tt=(\alpha_0,\alpha_\ell,\ell)=(\pi/2,0,1)$.}
%\label{fig:stereo}
%\end{figure}

 %In Figure~\ref{fig:sublevels}, we represent a cross section of the sets $\sublevel{P}(\ell_P)$ and $\sublevel{K\tp K}(\Delta^2)$ across the $\pe_1O\ve_1$ plane. It is possible to verify that the conditions~\eqref{eq:Hellip} are verified for the given controller parameters.
  %
%Figure~\ref{fig:flowfields}, depicts the flow field that determines the evolution of $x_1$, restricted to the flow set for two different values of the logic variable ${y_1}$. On the left, the value of the logic variable ${y_1}$ corresponds to the initial condition given in~\eqref{eq:initx} and it is such that a jump is triggered. The picture on the right of Figure~\ref{fig:flowfields} depicts the flow field after the jump. In this case, not only does $x_1$ belong to the flow set, but it will converge to $r_1$ along geodesics, as shown in Lemma~\ref{lem:geo}.
%

The experimental setup is akin to that of~\cite{cabecinhas_nonlinear_2014}. 
%and described next for completeness. The vehicle that is used for the experiments is the \emph{Blade 200 QX} quadrotor in Agility Mode, meaning that we provide angular velocity and thrust commands to the vehicle which are then relayed to the rotors by an internal controller which tracks the given commands. To measure the vehicle's position and orientation, we use the VICON motion capture system with 12 cameras (c.f.~\cite{VICON}), which provides measurements at a rate of 100 samples per second, with very high accuracy. The velocity of the vehicle is estimated from the position measurements. This experimental setup is illustrated in the diagram of Figure~\ref{fig:setup}.
%\begin{comment}
%
\begin{figure}
\centering
\psfragscanon
\psfrag{wb}[][]{$\wb$}
\psfrag{T}[][]{$u$}
\psfrag{p}[][]{$p$}
\psfrag{v}[][]{$v$}
\psfrag{R}[][]{$\Rmat$}
\psfrag{d}[][]{$\frac{d}{dt}$}
\psfrag{vicon}[][]{VICON}
\psfrag{blade}[][]{Blade 200 QX}
\psfrag{c}[][]{Controller}
\includegraphics[width=0.45\textwidth]{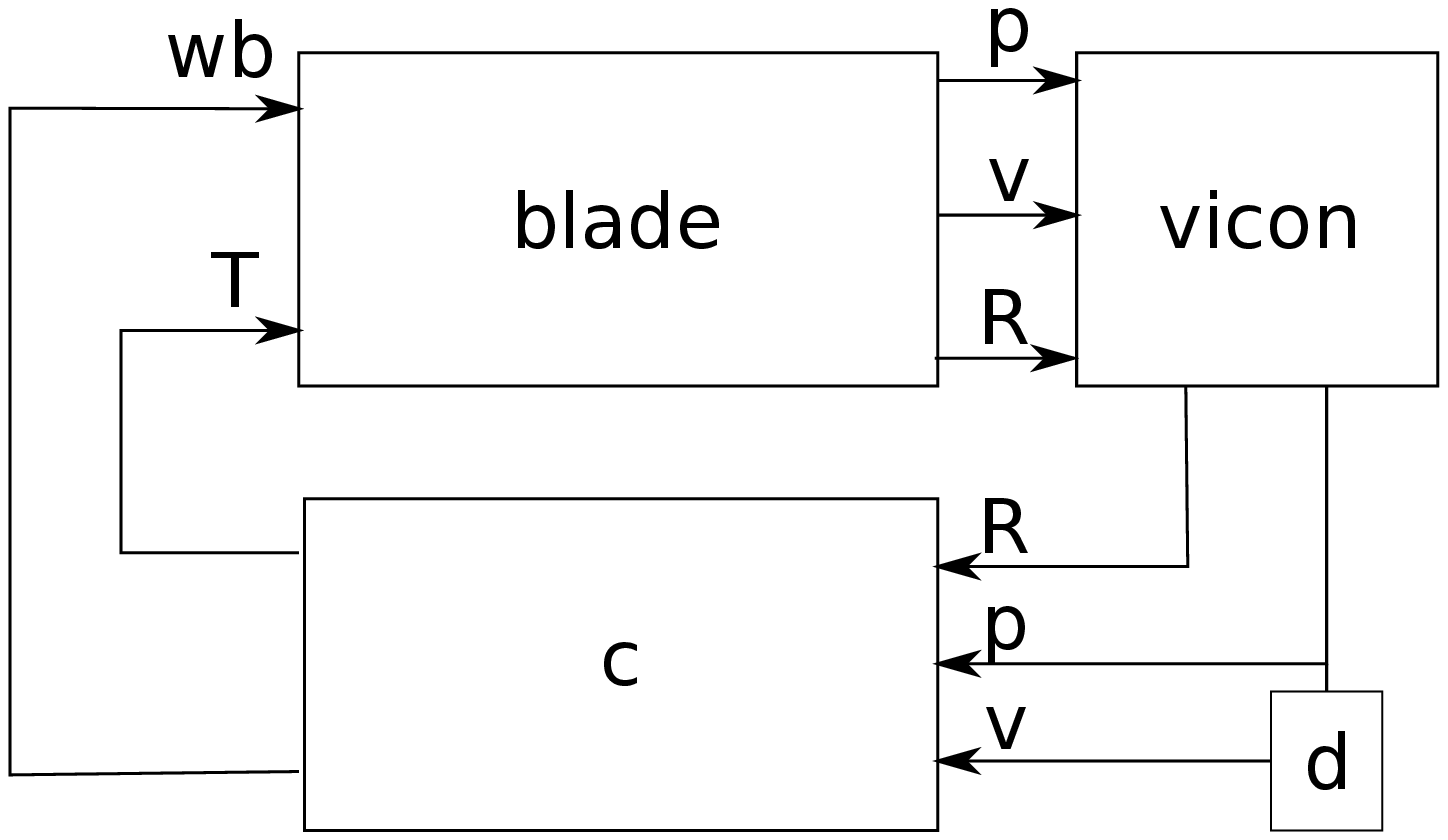}
\caption{Diagram representing the experimental setup.}
\label{fig:setup}
\end{figure}
%
%\end{comment}
To test the proposed controller under the given setup, we conducted a series of experiments which involved throwing the quadrotor up in the air and switching on the controller at the top of its trajectory using as a reference position that of the moment the controller was switched on. In the experiments, we set $k_p=0.05$ in order to compensate for the lag in angular velocity tracking. Alternatively, we could have reduced the controller gains of the position controller $K$ at the expense of controller performance. %In these experiments, we used the same controller gains that were chosen for the simulations in Section~\ref{sec:sim}, except for $k_p$, which was set to $0.05$ in order to compensate for the lag in angular velocity tracking. Alternatively, we could have reduced the controller gains of the position controller $K$ at the expense of controller performance. 
Figure~\ref{fig:exp} shows the results of one of the experimental runs.

The initial position error is zero because the reference position is the quadrotor  position when the controller is activated. Note that the initial velocity error is very high because the quadrotor is being thrown into the working area and the reference velocity is set to zero. It is possible to verify that the magnitude of the noise in the velocity measurements is very high in the first few seconds of the experiment. This is due to two main factors: since the quadrotor is being thrown into the working area upside down, there are a several reflective markers which disappear from the optical motion capture system field of view, thus reducing the accuracy in the position estimation. These imprecisions are compounded by the fact that the velocity is estimated from the position measurements by determining their rate of change. This noise has an effect in the determination of the desired thrust vector which induces controller switching at points during the first few seconds of the experiment. Nevertheless, the orientation error $V_1$ has an overall decrease in the first 2.5 s of the experiment and is kept near zero for the larger portion of the experiment which corresponds to zero orientation error. More importantly, the position and velocity converge to zero within the represented time span. {This controller can be used together with higher-level controllers to perform aggressive maneuvers as illustrated in~\cite{Casau2017}.}

\begin{figure*}
\centering
\input{exp.tex}
\includegraphics[width=\textwidth]{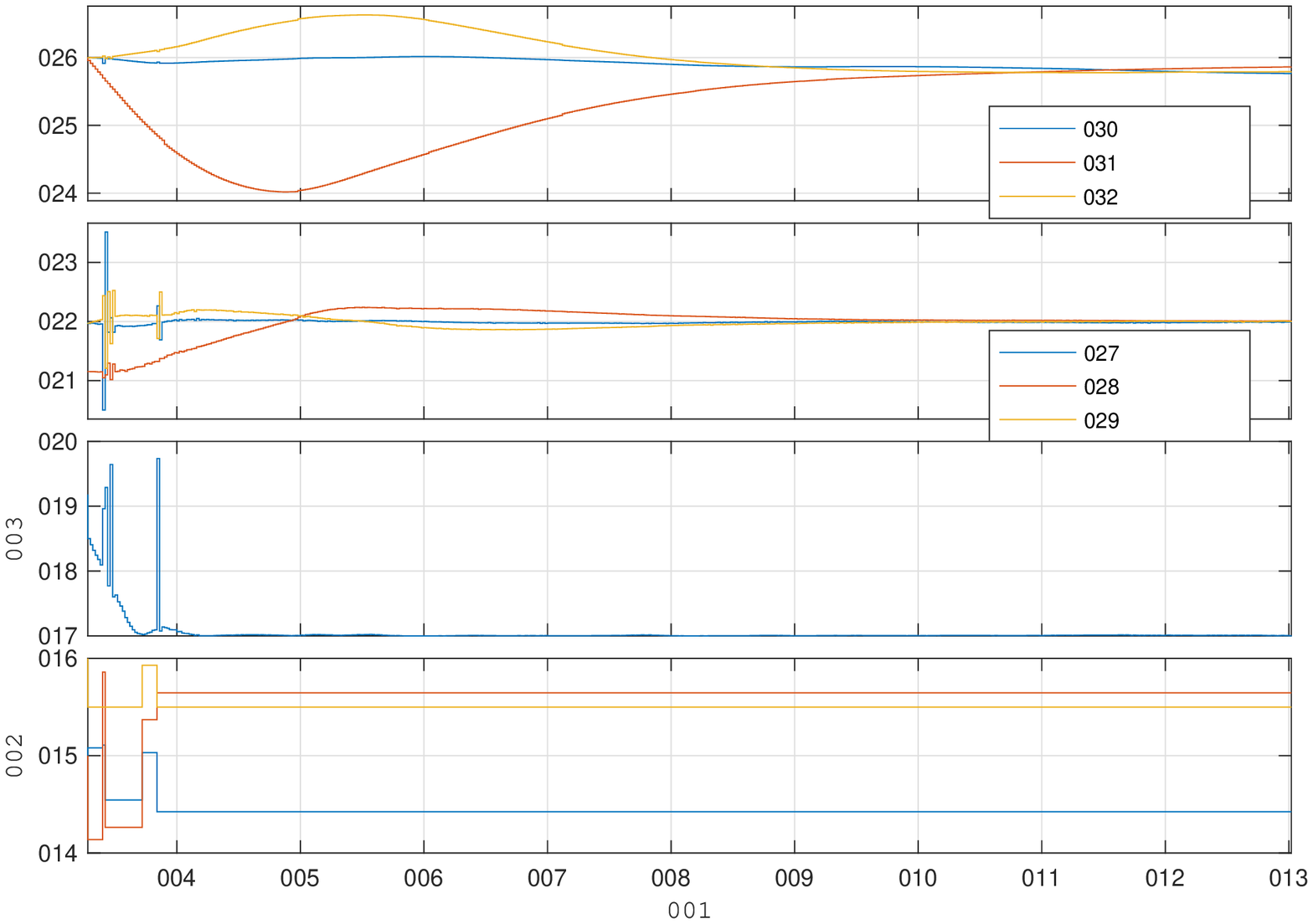}
\caption{Experimental results -- the quadrotor is thrown into the Vicon workspace and the controller is activated at the top of the trajectory.}
\label{fig:exp}
\hrulefill
\end{figure*}
}{}

\section{Conclusions}\label{sec:conclusions}
In this paper, we have demonstrated that the existence of synergistic potential functions on $\sphere n$ is a sufficient condition for the asymptotic stabilization of systems evolving on the $n$-dimensional sphere. Moreover, if these functions and their derivatives satisfy some additional bounds, it is possible to achieve global exponential stability. We provided a construction of synergistic potential functions which generates flows along geodesics upon switching. The proposed controller was then applied to trajectory tracking for a vectored thrust vehicle.

\bibliography{biblio}

\begin{thebibliography}{41}
\providecommand{\natexlab}[1]{#1}
\providecommand{\url}[1]{\texttt{#1}}
\expandafter\ifx\csname urlstyle\endcsname\relax
  \providecommand{\doi}[1]{doi: #1}\else
  \providecommand{\doi}{doi: \begingroup \urlstyle{rm}\Url}\fi

\bibitem[Augugliaro et~al.(2013)Augugliaro, Schoellig, and
  D'Andrea]{augugliaro_dance_2013}
F.~Augugliaro, A.~P. Schoellig, and R.~D'Andrea.
\newblock {Dance of the Flying Machines: Methods for Designing and Executing an
  Aerial Dance Choreography}.
\newblock \emph{IEEE Robotics {\&} Automation Magazine}, 20\penalty0
  (4):\penalty0 96--104, 2013.

\bibitem[Berkane and Tayebi(2015)]{berkane_design_2015}
S.~Berkane and A.~Tayebi.
\newblock {On the design of synergistic potential functions on SO(3)}.
\newblock In \emph{Proceedings of the 54th IEEE Conference on Decision and
  Control}, pages 270--275, 2015.

\bibitem[Berkane and Tayebi(2017)]{berkane_construction_tac2016}
S.~Berkane and A.~Tayebi.
\newblock {Construction of Synergistic Potential Functions on SO(3) with
  Application to Velocity-Free Hybrid Attitude Stabilization}.
\newblock \emph{IEEE Transactions on Automatic Control}, 62\penalty0 (1), 2017.

\bibitem[Berkane et~al.(2017)Berkane, Abdessameud, and
  Tayebi]{berkane_hybrid_2017}
S.~Berkane, A.~Abdessameud, and A.~Tayebi.
\newblock {Hybrid global exponential stabilization on SO(3)}.
\newblock \emph{Automatica}, 81:\penalty0 279--285, 2017.

\bibitem[Bhat and Bernstein(2000)]{bhat_topological_2000}
S.~P. Bhat and D.~S. Bernstein.
\newblock {A topological obstruction to continuous global stabilization of
  rotational motion and the unwinding phenomenon}.
\newblock \emph{Systems {\&} Control Letters}, 39\penalty0 (1):\penalty0
  63--70, 2000.

\bibitem[Brahmi et~al.(2019)Brahmi, Saad, Rahman, and
  Ochoa-Luna]{brahmi_cartesian_2017}
B.~Brahmi, M.~Saad, M.~H. Rahman, and C.~Ochoa-Luna.
\newblock {Cartesian Trajectory Tracking of a 7-DOF Exoskeleton Robot Based on
  Human Inverse Kinematics}.
\newblock \emph{IEEE Transactions on Systems, Man, and Cybernetics: Systems},
  49\penalty0 (3):\penalty0 600--611, 2019.

\bibitem[Burns and Gidea(2005)]{burns_differential_2005}
K.~Burns and M.~Gidea.
\newblock \emph{{Differential geometry and topology : with a view to dynamical
  systems}}.
\newblock CRC Press, 2005.

\bibitem[Casau et~al.(2015{\natexlab{a}})Casau, Mayhew, Sanfelice, and
  Silvestre]{Casau2015}
P.~Casau, C.~G. Mayhew, R.~G. Sanfelice, and C.~Silvestre.
\newblock {Global exponential stabilization on the n-dimensional sphere}.
\newblock In \emph{Proceedings of the 2015 American Control Conference (ACC)},
  pages 3218--3223, 2015{\natexlab{a}}.

\bibitem[Casau et~al.(2015{\natexlab{b}})Casau, Sanfelice, Cunha, Cabecinhas,
  and Silvestre]{casau_robust_2015}
P.~Casau, R.~Sanfelice, R.~Cunha, D.~Cabecinhas, and C.~Silvestre.
\newblock {Robust global trajectory tracking for a class of underactuated
  vehicles}.
\newblock \emph{Automatica}, 58:\penalty0 90--98, 2015{\natexlab{b}}.

\bibitem[Casau et~al.(2016)Casau, Mayhew, Sanfelice, and
  Silvestre]{casau_exponential_2016}
P.~Casau, C.~G. Mayhew, R.~G. Sanfelice, and C.~Silvestre.
\newblock {Exponential stabilization of a vectored-thrust vehicle using
  synergistic potential functions}.
\newblock In \emph{Proceedings of the 2016 American Control Conference}, pages
  6042--6047, 2016.

\bibitem[Casau et~al.(2017{\natexlab{a}})Casau, Mayhew, Sanfelice, and
  Silvestre]{Casau2017}
P.~Casau, C.~G. Mayhew, R.~G. Sanfelice, and C.~Silvestre.
\newblock {Exponential Stabilization of a Quadrotor Vehicle},
  2017{\natexlab{a}}.
\newblock URL \url{https://youtu.be/1DPKpxQPsMk}.

\bibitem[Casau et~al.(2017{\natexlab{b}})Casau, Sanfelice, and
  Silvestre]{casau_hybrid_2017}
P.~Casau, R.~G. Sanfelice, and C.~Silvestre.
\newblock {Hybrid Stabilization of Linear Systems with Reverse Polytopic Input
  Constraints}.
\newblock \emph{IEEE Transactions on Automatic Control}, 62\penalty0
  (12):\penalty0 6473 -- 6480, 2017{\natexlab{b}}.

\bibitem[Chaturvedi and McClamroch(2009)]{chaturvedi_asymptotic_2009}
N.~Chaturvedi and H.~McClamroch.
\newblock {Asymptotic Stabilization of the Inverted Equilibrium Manifold of the
  3-D Pendulum Using Non-Smooth Feedback}.
\newblock \emph{IEEE Transactions on Automatic Control}, 54\penalty0
  (11):\penalty0 2658--2662, 2009.

\bibitem[Chaturvedi et~al.(2011)Chaturvedi, Sanyal, and
  McClamroch]{chaturvedi_rigid-body_2011}
N.~Chaturvedi, A.~Sanyal, and N.~McClamroch.
\newblock {Rigid-Body Attitude Control}.
\newblock \emph{IEEE Control Systems}, 31\penalty0 (3):\penalty0 30--51, 2011.

\bibitem[Fink et~al.(2011)Fink, Michael, Kim, and Kumar]{fink_planning_2011}
J.~Fink, N.~Michael, S.~Kim, and V.~Kumar.
\newblock {Planning and control for cooperative manipulation and transportation
  with aerial robots}.
\newblock \emph{The International Journal of Robotics Research}, 30\penalty0
  (3):\penalty0 324--334, 2011.

\bibitem[Floreano and Wood(2015)]{floreano_science_2015}
D.~Floreano and R.~J. Wood.
\newblock {Science, technology and the future of small autonomous drones}.
\newblock \emph{Nature}, 521\penalty0 (7553):\penalty0 460--466, may 2015.

\bibitem[Goebel et~al.(2012)Goebel, Sanfelice, and Teel]{goebel_hybrid_2012}
R.~Goebel, R.~Sanfelice, and A.~Teel.
\newblock \emph{{Hybrid Dynamical Systems: Modeling, Stability, and
  Robustness}}.
\newblock Princeton University Press, 2012.

\bibitem[Hamel et~al.(2002)Hamel, Mahony, Lozano, and
  Ostrowski]{hamel_dynamic_2002}
T.~Hamel, R.~Mahony, R.~Lozano, and J.~Ostrowski.
\newblock {Dynamic Modelling and Configuration Stabilization for an X4-Flyer.}
\newblock \emph{IFAC Proceedings Volumes}, 35\penalty0 (1):\penalty0 217--222,
  2002.

\bibitem[Hua et~al.(2009)Hua, Hamel, Morin, and Samson]{hua_control_2009}
M.-D. Hua, T.~Hamel, P.~Morin, and C.~Samson.
\newblock {A Control Approach for Thrust-Propelled Underactuated Vehicles and
  its Application to VTOL Drones}.
\newblock \emph{IEEE Transactions on Automatic Control}, 54\penalty0
  (8):\penalty0 1837--1853, 2009.

\bibitem[Hua et~al.(2015)Hua, Hamel, Morin, and Samson]{hua_control_2015}
M.-D. Hua, T.~Hamel, P.~Morin, and C.~Samson.
\newblock {Control of VTOL vehicles with thrust-tilting augmentation}.
\newblock \emph{Automatica}, 52:\penalty0 1--7, 2015.

\bibitem[Jiang and Kumar(2013)]{jiang_inverse_2013}
Q.~Jiang and V.~Kumar.
\newblock {The Inverse Kinematics of Cooperative Transport With Multiple Aerial
  Robots}.
\newblock \emph{IEEE Transactions on Robotics}, 29\penalty0 (1):\penalty0
  136--145, feb 2013.

\bibitem[Lee(2015)]{lee_global_2015}
T.~Lee.
\newblock {Global Exponential Attitude Tracking Controls on SO(3)}.
\newblock \emph{IEEE Transactions on Automatic Control}, 60\penalty0
  (10):\penalty0 2837--2842, 2015.

\bibitem[Lee et~al.(2013)Lee, Leok, and McClamroch]{lee_nonlinear_2013}
T.~Lee, M.~Leok, and N.~H. McClamroch.
\newblock {Nonlinear Robust Tracking Control of a Quadrotor UAV on SE(3)}.
\newblock \emph{Asian Journal of Control}, 15\penalty0 (2):\penalty0 391--408,
  2013.

\bibitem[Liang et~al.(2018)Liang, Fang, Sun, and Lin]{liang_dynamics_2018}
X.~Liang, Y.~Fang, N.~Sun, and H.~Lin.
\newblock {Dynamics analysis and time-optimal motion planning for unmanned
  quadrotor transportation systems}.
\newblock \emph{Mechatronics}, 50:\penalty0 16--29, 2018.

\bibitem[Liz{\'{a}}rraga(2004)]{lizarraga_osbtructions_2004}
D.~A. Liz{\'{a}}rraga.
\newblock {Obstructions to the Existence of Universal Stabilizers for Smooth
  Control Systems}.
\newblock \emph{Mathematics of Control, Signals, and Systems (MCSS)},
  16\penalty0 (4):\penalty0 255--277, mar 2004.

\bibitem[Lupashin et~al.(2014)Lupashin, Hehn, Mueller, Schoellig, Sherback, and
  D'Andrea]{lupashin_platform_2014}
S.~Lupashin, M.~Hehn, M.~W. Mueller, A.~P. Schoellig, M.~Sherback, and
  R.~D'Andrea.
\newblock {A platform for aerial robotics research and demonstration: The
  Flying Machine Arena}.
\newblock \emph{Mechatronics}, 24\penalty0 (1):\penalty0 41--54, 2014.

\bibitem[Mahony et~al.(2012)Mahony, Kumar, and Corke]{mahony_multirotor_2012}
R.~Mahony, V.~Kumar, and P.~Corke.
\newblock {Multirotor Aerial Vehicles: Modeling, Estimation, and Control of
  Quadrotor}.
\newblock \emph{IEEE Robotics {\&} Automation Magazine}, 19\penalty0
  (3):\penalty0 20--32, 2012.

\bibitem[Markdahl et~al.(2018)Markdahl, Thunberg, and
  Gon{\c{c}}alves]{markdahl_almost_2018}
J.~Markdahl, J.~Thunberg, and J.~Gon{\c{c}}alves.
\newblock {Almost Global Consensus on the n-Sphere}.
\newblock \emph{IEEE Transactions on Automatic Control}, 63\penalty0
  (6):\penalty0 1664--1675, 2018.

\bibitem[Mayhew et~al.(2011{\natexlab{a}})Mayhew, Sanfelice, and
  Teel]{mayhew_quaternion-based_2011}
C.~Mayhew, R.~Sanfelice, and A.~Teel.
\newblock {Quaternion-Based Hybrid Control for Robust Global Attitude
  Tracking}.
\newblock \emph{IEEE Transactions on Automatic Control}, 56\penalty0
  (11):\penalty0 2555--2566, 2011{\natexlab{a}}.

\bibitem[Mayhew and Teel(2010)]{mayhew_global_2010}
C.~G. Mayhew and A.~R. Teel.
\newblock {Global asymptotic stabilization of the inverted equilibrium manifold
  of the 3-D pendulum by hybrid feedback}.
\newblock In \emph{Proceedings of the 49th IEEE Conference on Decision and
  Control}, pages 679--684. IEEE, 2010.

\bibitem[Mayhew and Teel(2011)]{mayhew_topological_2011}
C.~G. Mayhew and A.~R. Teel.
\newblock {On the topological structure of attraction basins for differential
  inclusions}.
\newblock \emph{Systems {\&} Control Letters}, 60\penalty0 (12):\penalty0
  1045--1050, 2011.

\bibitem[Mayhew and Teel(2013{\natexlab{a}})]{mayhew_global_2013}
C.~G. Mayhew and A.~R. Teel.
\newblock {Global stabilization of spherical orientation by synergistic hybrid
  feedback with application to reduced-attitude tracking for rigid bodies}.
\newblock \emph{Automatica}, 49\penalty0 (7):\penalty0 1945--1957,
  2013{\natexlab{a}}.

\bibitem[Mayhew and Teel(2013{\natexlab{b}})]{mayhew_synergistic_2013}
C.~G. Mayhew and A.~R. Teel.
\newblock {Synergistic hybrid feedback for global rigid-body attitude tracking
  on SO(3)}.
\newblock \emph{IEEE Transactions on Automatic Control}, 58\penalty0
  (11):\penalty0 2730--2742, 2013{\natexlab{b}}.

\bibitem[Mayhew et~al.(2011{\natexlab{b}})Mayhew, Sanfelice, and
  Teel]{Mayhew2011}
C.~G. Mayhew, R.~G. Sanfelice, and A.~R. Teel.
\newblock {Synergistic Lyapunov functions and backstepping hybrid feedbacks}.
\newblock In \emph{Proceedings of the 2011 American Control Conference}, pages
  3203--3208, 2011{\natexlab{b}}.

\bibitem[Mayhew et~al.(2012)Mayhew, Sanfelice, Sheng, Arcak, and
  Teel]{mayhew_quaternion-based_2012}
C.~G. Mayhew, R.~G. Sanfelice, J.~Sheng, M.~Arcak, and A.~R. Teel.
\newblock {Quaternion-Based Hybrid Feedback for Robust Global Attitude
  Synchronization}.
\newblock \emph{IEEE Transactions on Automatic Control}, 57\penalty0
  (8):\penalty0 2122--2127, aug 2012.

\bibitem[Mellinger et~al.(2012)Mellinger, Michael, and
  Kumar]{mellinger_trajectory_2012}
D.~Mellinger, N.~Michael, and V.~Kumar.
\newblock {Trajectory generation and control for precise aggressive maneuvers
  with quadrotors}.
\newblock \emph{The International Journal of Robotics Research}, 31\penalty0
  (5):\penalty0 664--674, 2012.

\bibitem[Petersen(2006)]{petersen_riemannian_2006}
P.~Petersen.
\newblock \emph{{Riemannian Geometry}}, volume 171 of \emph{Graduate Texts in
  Mathematics}.
\newblock Springer New York, 2006.

\bibitem[Saberi et~al.(2012)Saberi, Stoorvogel, and
  Sannuti]{saberi_internal_2012}
A.~Saberi, A.~Stoorvogel, and P.~Sannuti.
\newblock \emph{{Internal and External Stabilization of Linear Systems with
  Constraints}}.
\newblock Systems {\&} Control: Foundations {\&} Applications. Birkh{\"{a}}user
  Boston, Boston, MA, 2012.

\bibitem[Sundaram(1996)]{Sundaram1996}
R.~Sundaram.
\newblock \emph{{A First Course in Optimization Theory}}.
\newblock Cambridge University Press, New York, USA, 20th edition, 1996.

\bibitem[Triantafyllou et~al.(2018)Triantafyllou, Rovithakis, and
  Doulgeri]{triantafyllou_constrained_2018}
P.~Triantafyllou, G.~Rovithakis, and Z.~Doulgeri.
\newblock {Constrained visual servoing under uncertain dynamics}.
\newblock \emph{International Journal of Control}, 2018.

\bibitem[Zhao and Zelazo(2016)]{zhao_bearing_2016}
S.~Zhao and D.~Zelazo.
\newblock {Bearing Rigidity and Almost Global Bearing-Only Formation
  Stabilization}.
\newblock \emph{IEEE Transactions on Automatic Control}, 61\penalty0
  (5):\penalty0 1255--1268, 2016.

\end{thebibliography}

\appendix

\section{Auxiliary results}
\begin{apro}\label{prop:muIsNice}
Let $r\in\sphere n$ and let $\yset$ be a compact set. Given $V\in\spf[r]{\yset}$, the following holds:
\begin{enumerate}
\item \label{item:Vpd} $r\in \crit(V^y)$ for each $y\in\yset$;
\item \label{item:muC}The function~\eqref{eq:mu} is continuous and the map~\eqref{eq:yminV} is outer semicontinuous.
\end{enumerate}
\end{apro}

\begin{pf}
\IfFull{%
{Since $V$ is assumed to be continuously differentiable and $V^y$ is positive definite relative to $r$, this implies that $r\in\crit V^y$ for each $y\in\yset$. 
From the continuity of $V$, we have that~\eqref{eq:mu} is continuous if and only if $\minVy(x)\ceq\min_{\bar y\in \yset}V(x,\bar y)$ is continuous. Let $\{x_i\}_{i\in\N}$ denote a sequence that converges to some $x\in\sphere n$ and let $\{y_i\}_{i\in\N}$ be such that $y_i\in\arg\min_{\bar y\in \yset} V(x_i,\bar y).$ Since $\yset$ is compact, there is at least one convergent subsequence of $\{y_i\}_{i\in\N}$. For every convergent subsequence, denoted by $\{y_{k(i)}\}_{i\in\N}$, with accumulation point $y_k\in \yset$, we conclude that $V(x_{k(i)},y_{k(i)})$ converges to $V(x,y_k)$ because $V$ is continuous. Suppose that there exists $z\in\yset$ such that $V(x,z)<V(x,y_k)$ and suppose that $\{z_i\}_{i\in\N}$ is a sequence that converges to $z$. By the continuity of $V$ and by the assumption that $V(x,z)<V(x,y_k)$, there exists $I\in\N$ such that $V(x_{k(i)},z_{k(i)})<V(x_{k(i)},y_{k(i)})$ for each $i\geq I$. However, this is a contradiction because $y_{k(i)}$ is the minimizer for $V(x_{k(i)},y_{k(i)})$. We conclude that $\lim_{i\rightarrow\infty}\minVy(x_i)=\minVy(x)$, thus $\minVy(x)$ is continuous. Also, for every sequence $\{x_i\}_{i\in\N}$ convergent to $x\in\sphere n$, and for each sequence $\{y_i\}_{i\in\N}$ convergent to $y\in\yset$ such that $y_i\in\yminV(x_i)$ for each $i\in\N$, we have that $y\in\yminV(x)$ which proves that $\yminV(x)$ is outer semicontinuous.}
}{This result follows from the application of the Maximum Theorem in~\cite{Sundaram1996} and computations similar to~\cite[Proposition~1]{mayhew_global_2013}.}
\qed
\end{pf}

\begin{alem}\label{lem:V}
The function $V : \Vdom \to \R{}$ satisfies
\begin{align}
\label{eqn:argminV}
\begin{split}
\argmin_{(x,y) \in \Vdom} V(x,y) &= V^{-1}(0) = \{(r,y) \in \Vdom\}
\end{split} \\
\label{eqn:argmaxV}
\begin{split}
\argmax\limits_{(x,y) \in \Vdom} V(x,y) &= V^{-1}(1) = \{(x,x) \in \Vdom\}
\end{split}
\end{align}
Moreover, $V(x,y)$ is positive definite on $\Vdom$ relative to $\{(r,y) \in \Vdom\}$ and for each $x \in \sphere n$,
\begin{align}\label{eqn:argminyV}
\argmin_{y \in \sphere n} V(x,y) &= \begin{cases} -x & \text{if } x \neq r \\
\sphere n\minus\{r\} & \text{if } x = r
\end{cases} \\
\label{eqn:minyV}
\min_{y \in \sphere n} V(x,y) &= \frac{1-r\tp  x}{2k+1 - r\tp  x}.
\end{align}
\end{alem}
\begin{pf}
Since $\height{r}(x)\geq 0$ for all $x\in\sphere n$ and $\height{r}(x)=0$ if and only if $x=r$, it follows that $h_{r}(x) + kh_{y}(x) > 0$ on $\Vdom$. Setting $V(x,y) \leq 0$, we find that $h_{r}(x) \leq 0$, which can only be satisfied (with equality) when $x = r$. That is, $V$ attains its minimum value of zero on the set $\{(r,y) \in \Vdom\}$. Similarly, $V(x,y) \geq 1$ if and only if $h_{y}(x) \leq 0$, which is again satisfied (with equality) only when $x = y$ and thus, $V$ attains its maximum value of one on the set $\{(x,x) \in \Vdom\}$.
To prove \eqref{eqn:argminyV} and \eqref{eqn:minyV}, we note---from our previous observations---that
\begin{equation}\label{eqn:minyV2}
\begin{split}
\min_{\bar y \in \sphere n} V(x,\bar y) %& = \min_{y \in \sphere n} \frac{1-r\tp  x}{1-r\tp  x + k\left(1-y\tp  x\right)} \\
&= \frac{1-r\tp  x}{1-r\tp  x + k\max_{\bar y \in \sphere n} (1-\bar y\tp  x)}.
\end{split}
\end{equation}
Since, for each $x \in \sphere n$, $1-y\tp  x$ attains its maximal value of two at $y = -x$, this choice minimizes $y \mapsto V(x,y)$ for each $x \in \sphere n\setminus\{r\}$; however, when $x = r$, $V(x,y) = 0$ for any $y \in \sphere n\setminus\{r\}$, thus proving \eqref{eqn:argminyV}. Eq.~\eqref{eqn:minyV} follows from evaluating $V(x,y)$ with $y = -x$.
\qed\end{pf}

\begin{acor}\label{cor:critV}
Given $y \in \sphere n$, define $V^y: \sphere n \to \R{}$ for each $x \in \sphere n$ as $V^y(x) = V(x,y)$, with $V$ given by~\eqref{eqn:V}. Then,
\begin{equation}\label{eqn:critV}
\crit V^y= \begin{cases}
\{r,y\} & \text{if } r \neq y \\
\sphere n & \text{otherwise}.
\end{cases}
\end{equation}
\end{acor}
\begin{pf}
By definition, $x \in \crit V^y$ if and only if $\PTSn(x)\grad V^y(x) = 0$, or equivalently, $\left|\PTSn(x)\grad V^y(x)\right| = 0$.  Noting that $V^y(x) = V(x,y)$ for each $x \in \sphere n$, it follows that $\grad V^y(x) = \gradx V(x,y)$ for each $x \in \sphere n$. By \eqref{eqn:normVgradx} of Lemma~\ref{lem:gradV}, it follows that $x \in \crit V^y$ if and only if 
\begin{equation}\label{eqn:critVpoints}
V^y(x)(1-V^y(x))(1-r\tp  y) = 0.
\end{equation}
Clearly, \eqref{eqn:critVpoints} is satisfied in three cases: $V^y(x) = 0$, $1-V^y(x) = 0$, or $1-r\tp  y = 0$. 
When $1-r\tp  y = 0$, or equivalently, when $r = y$, it follows that every point in $\sphere n$ is a critical point, since \eqref{eqn:critVpoints} is satisfied for every $x \in \sphere n$. In fact, when $r = y$, $V^y(x) = 1/(1+k)$, so that for all $x \in \sphere n$, $\grad V^y(x) = 0$ and obviously $\PTSn(x)\grad V^y(x) = 0$. We now examine the remaining cases.
When $V^y(x) = 0$, it follows from the definition of $V^y(x) = V(x,y)$ that $1-r\tp  x = 0$, or equivalently, $x = r$. If $V^y(x) = 1$, a short calculation yields $1-y\tp  x = 0$, or equivalently, $x = y$. Thus, when $r \neq y$, $\crit V^y= \{r,y\}$.
\qed\end{pf}
\begin{alem}\label{lem:globaldenombnd}
The following holds
\begin{equation}\label{eqn:globaldenombnd}
\begin{multlined}
0< 1+k-\sqrt{1+2k\yrbnd+k^2} \leq 1-r\tp   x + k(1-y\tp   x) \\ \leq 1+k+\sqrt{1+2k\yrbnd+k^2}.
\end{multlined}
\end{equation}
for all $(x,y) \in \sphere n\x\yset$.
\end{alem}
\begin{pf}
The upper and lower bounds on~\eqref{eqn:globaldenombnd} follow from the solution to the optimization problem $\min/\max\{J(x,y):1-y\tp   y = 0,1-x\tp   x = 0,r\tp   y - \yrbnd = 0\}$
with $J(x,y) = 1-r\tp   x + k(1-y\tp   x)$ for each $(x,y) \in \sphere n\x\yset$, by means of Lagrange multipliers.
\IfFull{%
To prove this bound, we solve the following optimization problem:
\begin{equation}\label{eqn:minmaxdenomV}
\begin{aligned}[t]
\text{min/max} &\quad J(x,y) \\
\text{subject to} 
&\quad 1-y\tp   y = 0 \\ 
&\quad 1-x\tp   x = 0 \\
&\quad r\tp   y - \yrbnd = 0,
\end{aligned}
\end{equation}
where $J(x,y) = 1-r\tp   x + k(1-y\tp   x)$ for each $(x,y) \in \sphere n\x\yset$.
We proceed by constructing the Lagrangian, $L: \sphere n\x\yset \times \R{}\times \R{} \times \R{} \to \R{}$ for each $(x,y,\lambda_{1},\lambda_{2},\beta) \in \sphere n\x\yset \times \R{} \times \R{} \times \R{}$ as
\begin{equation}\label{eqn:lagrangian2}
\begin{aligned}
L(x,y,\lambda_{1},\lambda_{2},\beta) &= J(x,y) + \frac{\lambda_{1}}{2}(1-y\tp   y)\\
& + \frac{\lambda_{2}}{2}(1-x\tp   x) + \beta(r\tp   y - \yrbnd).
\end{aligned}
\end{equation}
According to the KKT conditions, each solution to \eqref{eqn:minmaxdenomV} must satisfy $\grad L = 0$, $\beta \geq 0$, equivalently
\begin{align}
\label{eqn:stationary1}
\lambda_{1}y &= \beta r - kx \\
\label{eqn:stationary2}
\lambda_{2}x &= -r - ky,
\end{align}
in addition to $\beta \geq 0$ and 
\begin{equation}\label{eqn:slackness}
\beta(r\tp   y - \yrbnd) = 0.
\end{equation}
We break the analysis into cases. Assuming that $\beta = 0$, it follows from \eqref{eqn:stationary1} and \eqref{eqn:stationary2} that $x = -y = r$, where $J(r,-r) = 2k$ or $x = y = -r$, where $J(-r,-r) = 2$.

Assuming that $\beta \neq 0$, it follows from \eqref{eqn:slackness} that $r\tp   y = \yrbnd$. Then, by multiplying \eqref{eqn:stationary2} on the left by $x\tp  $, it follows that 
\begin{equation}
\lambda_{2} = -r\tp   x -ky\tp   x
\end{equation}
so that $J(x,y) = 1+k+\lambda_{2}$. In the same direction, by taking the norm of both sides of \eqref{eqn:stationary1} and applying the fact that $r\tp   y = \yrbnd$, it follows that
\begin{equation}
\lambda_{2}^{2} = 1+2k\yrbnd+k^{2},
\end{equation}
so that $J(x,y) = 1+k\pm\sqrt{1+2k\yrbnd+k^{2}}$.

Now that we have all possible values of $J(x,y)$ that solve \eqref{eqn:minmaxdenomV}, we evaluate to find the minimum and maximum of $J$. Clearly, $1+k-\sqrt{1+2k\yrbnd+k^{2}} < 1+k+\sqrt{1+2k\yrbnd+k^{2}}$. Moreover, it is obvious that $1\leq\sqrt{1+2k\yrbnd+k^{2}}$ and $k\leq\sqrt{1+2k\yrbnd+k^{2}}$, thus $\max\{2,2k\} \leq 1+k+\sqrt{1+2k\yrbnd+k^{2}}$.
}{}
\qed\end{pf}

\end{document}